\documentclass[pra,twocolumn,showpacs,superscriptaddress,floatfix, nofootinbib]{revtex4-1}
\usepackage[caption=false]{subfig}
\usepackage[tbtags]{amsmath}
\usepackage{graphicx}
\usepackage{epsfig}
\usepackage{amssymb,mathrsfs,esint, amsbsy}
\usepackage{xcolor}
\usepackage{soul}
\usepackage{gensymb}
\usepackage[bookmarks=true,
   colorlinks=true,
   linkcolor=blue,
   urlcolor=blue,
   citecolor=blue,
   bookmarks=true,
   hyperindex=true
]{hyperref}
\usepackage{natbib}
\usepackage{relsize}
\usepackage{float}
\usepackage{dsfont}

\usepackage{tikz, wasysym}
\usepackage{bm}

\newcommand{\be}{\begin{equation}}
\newcommand{\ee}{\end{equation}}

\newcommand{\beq}{\begin{eqnarray}}
\newcommand{\eeq}{\end{eqnarray}}

\def\H1{\widehat{H}_1}

\newcommand{\ket}[1]{\left| #1 \right>}
\newcommand{\bra}[1]{\left< #1 \right|}

\newcommand{\bs}{\boldsymbol}

\begin{document}

\title{ Disordered impenetrable two-component fermions in one dimension}

\author{D.\,V. Kurlov}
\affiliation{Russian Quantum Center, Skolkovo, Moscow 121205, Russia}
\affiliation{National University of Science and Technology ``MISIS”, Moscow 119049, Russia}

\author{M.\,S. Bahovadinov}
\affiliation{Russian Quantum Center, Skolkovo, Moscow 121205, Russia}
\affiliation{  Physics Department, National Research University Higher School of Economics, Moscow, 101000, Russia}

\author{S.\,I. Matveenko}
\affiliation{L.\,D. Landau Institute for Theoretical Physics, Chernogolovka, Moscow region 142432, Russia}
\affiliation{Russian Quantum Center, Skolkovo, Moscow 143025, Russia}

\author{A.\,K.~Fedorov}
\affiliation{Russian Quantum Center, Skolkovo, Moscow 121205, Russia}
\affiliation{National University of Science and Technology ``MISIS”, Moscow 119049, Russia}
\affiliation{Schaffhausen Institute of Technology, Schaffhausen 8200, Switzerland}

\author{V.~Gritsev}
\affiliation{Institute for Theoretical Physics Amsterdam, Universiteit van Amsterdam, Amsterdam 1098 XH, The Netherlands}
\affiliation{Russian Quantum Center, Skolkovo, Moscow 121205, Russia}

\author{B.\,L. Altshuler}
\affiliation{Physics Department, Columbia University, 538 West 120th Street, New York, New York 10027, USA}
\affiliation{Russian Quantum Center, Skolkovo, Moscow 121205, Russia}

\author{G.\,V. Shlyapnikov}
\affiliation{Russian Quantum Center, Skolkovo, Moscow 121205, Russia}
\affiliation{Moscow Institute of Physics and Technology, Dolgoprudny, Moscow Region 141700, Russia}
\affiliation{Universit\'e Paris-Saclay, CNRS, LPTMS, 91405 Orsay, France}
\affiliation{Van der Waals-Zeeman Institute, Institute of Physics, University of Amsterdam,Science Park 904, 1098 XH Amsterdam, The Netherlands}

\begin{abstract}
We study the one-dimensional Hubbard model for two-component fermions with infinitely strong on-site repulsion ($t\!-\!0$ model) in the presence of disorder. Our analytical treatment demonstrates that the type of disorder drastically changes the nature of the emerging phases. The case of spin-independent disorder can be treated as a single-particle problem with Anderson localization. On the contrary, recent numerical findings show that spin-dependent disorder, which can be realized as a random magnetic field, leads to the many-body localization-delocalization transition.
We find an explicit analytic expression for the matrix elements of the random magnetic field between the eigenstates of the  $t-0$ model with potential disorder on a finite lattice. Analysis of the matrix elements supports the existence of the many-body localization-delocalization transition in this system and provides an extended physical picture of the random magnetic field. 
\end{abstract}

\maketitle

\section{Introduction}

Anderson localization (AL) describing the behaviour of non-interacting quantum particles in disorder is a cornerstone concept in condensed matter physics~\cite{Evers2008}. A conceptual extension of this effect in the presence of interparticle interactions is known as many-body localization (MBL)~\cite{Basko2006, Oganesyan2007}. In the vast majority of settings, the MBL problem is extremely difficult for analytical treatments. Thus, to a large extent the progress in the field is driven by numerics. In particular, vanishing steady transport, absence of thermalization, and area-law scaling of entanglement entropy in the MBL phase were demonstrated numericaly~\cite{Berkelbach2010, Barisic2016, YBLev2015, Herbrych2017, Luitz2016a, Serbyn2013, De_Luca_2013, Nayak_2013} (for review, see e.g.~\cite{Luitz2017, Abanin2017, Alet2018} and references therein). Traditional setups for analyzing MBL transition are one-dimensional (1D) models, such as the Heisenberg XXZ spin chain and the Fermi-Hubbard model~\cite{Luitz2015, Burin2015, Luitz2016, Kudo2018, Bahovadinov2022}. On the one hand, these systems are paradigmatically important for condensed matter. On the other hand, they can be in some limits treated analytically. In addition, these 1D systems can be realized using controllable ensembles of neutral atoms in optical potentials~\cite{Bloch2008, Guan2013, Mazurenko2017}, trapped ions~\cite{Bermudez2010}, and superconducting circuits~\cite{Barends2015, LasHeras2015}. 

The disordered Fermi-Hubbard model has been a subject of intensive research in the recent years~\cite{Mondaini2015, Bonca2017, Lemut2017, Kozarzewski2018, Zakrzewski2018, Kozarzewski2019, Iadecola2019, Krause2021}. An interest to this model is related to its rich physics, which is caused by various types of disorder (with respect to spin or charge), range of the strength of the interparticle interaction, and types of boundary conditions. Additional motivation comes from the recent experimental realization of the disordered Fermi-Hubbard chain in a system of cold neutral atoms~\cite{Schreiber2015, Scherg2021}. 

Recent numerical results suggest that a sufficiently strong random potential localizes the charge sector of the 1D Fermi-Hubbard chain, whereas the spin sector remains delocalized~\cite{Prelovsek2016, Protopopov2017, Protopopov2019, Sroda2019}. It was also reported that in a sufficiently strong random magnetic field spin excitations are localised whereas states in the charge sector remain extended~\cite{LeipnerJohns2019}. Recently, the disordered Fermi-Hubbard model for two-component fermions with infinitely strong on-site repulsion (the $t\!-\!0$ model) was investigated numerically~\cite{Bahovadinov2022}. It was demonstrated that in the spin-independent (potential) disorder the states exhibit AL. On the contrary, in a sufficiently weak spin-dependent disorder (random magnetic field) the states remain ergodic and the system undergoes the MBL transition at strong disorder. Reenterant AL-MBL for two-component disorder was also demonstrated~\cite{Bahovadinov2022}.

In this work, we provide analytical arguments to support these numerical findings. We employ factorization of the charge and spin degrees of freedom, which allows one to eliminate the spin degrees of freedom for both open and periodic boundary conditions. We then analytically study the system in two regimes: (i) potential disorder, which drives the corresponding single-particle system to AL and (ii) random magnetic field that causes possible localization-delocalization (MBL) transitions.

The paper is organized as follows. In Sec.~\ref{S:1D_Hubbard_overview}, we briefly overview the 1D Hubbard model with the focus on its strong-coupling regime, $U\to +\infty$. Then, in Sec.~\ref{S:t0_model_mapping} we discuss in detail the mapping of the $t\!-\!0$ model onto free fermions. 
In Sec.~\ref{S:Hubbard_disorder}, we study the effect of disorder in the limit of infinite inter-component on-site interaction. We demonstrate that in the case of spin-independent disorder the physical picture is drastically different from that in the case of spin-dependent disorder. 
In the former case the disordered $t-0$ Hamiltonian can be exactly mapped onto the tight-binding model of free spinless fermions with onsite disorder and quasi-periodic boundary conditions, which is well-known to exhibit AL. However, in the case of random magnetic field such mapping is no longer possible. We analytically calculate the matrix elements of the random magnetic field in the eigenbasis of the $t-0$ Hamiltonian with potential disorder. We show that the random magnetic field strongly couples the localized single-particle states, which may lead to the delocalization transition.
Finally, in Sec.~\ref{S:Conclusions} we summarize our results and conclude. 

\section{Clean 1D Hubbard model and its strong coupling limit} \label{S:1D_Hubbard_overview}

\subsection{1D Hubbard Hamiltonian}
The Hamiltonian of the 1D spin-$1/2$ Fermi-Hubbard model in a lattice of $L$ sites is given by 
\begin{multline} \label{Hubbard_H_gen}
	H_{\text{FH}}  = - t \sum_{j=1}^{L}  \sum_{\sigma=\uparrow,\downarrow} \left( c_{j, \sigma}^{\dag} c_{j + 1, \sigma} + \text{H.c.}  \right) \\
	 + U \sum_{j=1}^L \left( n_{j,\uparrow} - \frac{1}{2} \right) \left( n_{j,\downarrow} - \frac{1}{2} \right),
\end{multline}
where the operator $c_{j, \sigma}$ ($c_{j, \sigma}^{\dag}$) annihilates (creates) a fermion in the spin-$\sigma$ state at the~$j$th site of the lattice,~$n_{j,\sigma} = c^{\dag}_{j, \sigma} c_{j, \sigma}$ is the number operator, $t$~is the nearest-neighbor hopping amplitude, and~$U$ is the strength of the on-site interaction between different spin components. Fermionic creation and annihilation operators satisfy the canonical anticommutation relations~$\{ c_{j, \sigma}, c_{k, \sigma^{\prime}}^{\dag} \} = \delta_{j,k} \delta_{\sigma, \sigma^{\prime}}$ and  $\{ c_{j, \sigma}, c_{k, \sigma^{\prime}} \} = \{ c_{j, \sigma}^{\dag}, c_{k, \sigma^{\prime}}^{\dag} \} = 0$. Hereinafter we impose periodic boundary conditions, $c_{L+1,\sigma} = c_{1, \sigma}$, so that the Hamiltonian~(\ref{Hubbard_H_gen}) is translation invariant. 
For finite~$U$, the local Hilbert space~${\cal H}_j$ is four-dimensional and is spanned by the following states:
\be \label{4d_local_Hilbert_space}
	{\cal H}_j = \left\{ \ket{0}_j, \; c_{j,\uparrow}^{\dag}\ket{0}_j, \; c_{j,\downarrow}^{\dag}\ket{0}_j , \;  c_{j,\uparrow}^{\dag}c_{j,\downarrow}^{\dag}\ket{0}_j \right\},
\ee
so that the total Hilbert space~$\otimes_{j=1}^L{\cal H}_j$ is~$4^L$-dimensional.

Over the last decades, the Hamiltonian~(\ref{Hubbard_H_gen}) has been extensively studied in various contexts and there is a vast literature on its rich physics (see e.g. Ref.~\cite{Hubbard_book} for review). The model is exactly solvable by the (nested) Bethe ansatz for arbitrary~$U$, which was first shown in the pioneering work of Lieb and Wu~\cite{Lieb1968}. Alternative solution of the Hubbard model is provided by the algebraic Bethe ansatz, which relies on the so-called $R$-matrix that completely determines the model. For the Hubbard model~(\ref{Hubbard_H_gen}) the $R$-matrix was found by Shastry~\cite{Shastry1986}. 
 
 \subsection{1D Hubbard model in the limit of $U\to + \infty$} \label{S:t-0_model_BA}

In the limit of {\it infinitely} strong repulsion, $U\to + \infty$, the Hamiltonian (\ref{Hubbard_H_gen}) reduces to the so-called $t\!-\!0$ model described by the Hamiltonian~\cite{Hubbard_book}
\be \label{t_0_Hamiltonian}
	H_{t-0} = - t \; {\mathsf P} \sum_{j=1}^{L} \sum_{\sigma=\uparrow,\downarrow} \left( c_{j,\sigma}^{\dag} c_{j+1, \sigma} + \text{H.c.} \right) {\mathsf P},
\ee
where the Hermitian operator
\be \label{strong_coupling_P}
	{\mathsf P} = \prod_{j=1}^L\left( 1 - n_{j, \uparrow} n_{j, \downarrow} \right) 
\ee
satisfies~${\mathsf P}^2 ={\mathsf P}$ and projects out the states with doubly occupied sites. Thus, the physical Hilbert space becomes~$3^L$-dimensional, since the state~$c_{j,\uparrow}^{\dag} c_{j,\downarrow}^{\dag} \ket{0}_j$ in Eq.~(\ref{4d_local_Hilbert_space}) is separated from the rest by an infinite energy. 
Note that the $t\!-\!0$ model is non-trivial only away from half-filling.
By transferring the left projector~${\mathsf P}$ through the sums over $j$ and $\sigma$ to the right in Eq.~(\ref{t_0_Hamiltonian}), one can rewrite the $t\!-\!0$ Hamiltonian as
\begin{multline} \label{t_0_Hamiltonian_v2}
	H_{t-0} = - t \sum_{j=1}^L \sum_{\sigma = \uparrow, \downarrow} \Bigl(  c_{j,\sigma}^{\dag} c_{j+1, \sigma}\left( 1 - n_j \right) \\
	 + c_{j + 1,\sigma}^{\dag} c_{j, \sigma}\left( 1 - n_{j+1} \right) \Bigr) {\mathsf P},
\end{multline}
where~$n_j = n_{j,\uparrow} + n_{j,\downarrow}$, and the projector~${\mathsf P}$ in Eq.~(\ref{t_0_Hamiltonian_v2}) can be safely omitted and restored at will since the Hamiltonian~(\ref{t_0_Hamiltonian_v2}) acts on the projected Hilbert space~${\mathsf P}\otimes_{j=1}^L{\cal H}_j$.  

The $t-0$~model in Eq.~(\ref{t_0_Hamiltonian_v2}) possesses all conservation laws of the Hubbard Hamiltonian~(\ref{Hubbard_H_gen}) with a finite~$U$, plus an additional one. Indeed, since the doubly occupied states are not allowed in the $U \to + \infty$ limit, particles can not exchange their positions. As a result, the {\it spin pattern}, or ordering, is conserved. For open boundary conditions, the spin pattern is frozen and the spin degrees of freedom are completely decoupled from the charge sector. On the other hand, for periodic boundary conditions the situation is more subtle. In this case, because of the hopping between the boundary sites, the spin pattern can change, but only in a cyclic way. This cyclic change in the spin pattern results in the coupling between the charge and spin degrees of freedom.

In the limit of infinite $U$ the Bethe ansatz solution of the Hubbard model~(\ref{Hubbard_H_gen}) is drastically simplified, since all spin configurations become degenerate~\cite{Hubbard_book}. Because of this degeneracy, in the $t-0$ model the spin part of an eigenstate can be an {\it arbitrary} linear combination of those for the Hubbard model in the limit~$U\to +\infty$. In the case of periodic boundary conditions the only requirement is that the spin part of an eigenstate is also an eigenvector of the cyclic shift operator. As was shown in Refs.~\cite{Mielke1991, Izergin1997, Abarenkova1997, Izergin1998a, Izergin1998b}, it is convenient to take the eigenstates of the isotropic XY spin chain (also called the XX or XX$0$ model) for the spin part of the eigenstates of the $t\!-\!0$ model. The reason is that the XX model is non-interacting, so that its eigenstates are simple to deal with. This is in stark contrast with the case of an arbitrarily large but finite $U$. There, the spin sector of the Hubbard Hamiltonian can be effectively described by the XXX model, whose eigenstates are much more complicated since they correspond to {\it interacting} spin excitations~\cite{Hubbard_book, Ogata1990}.

Thus, in the sector with fixed numbers of spin-up and spin-down particles ($N_{\uparrow}$ and $N_{\downarrow}$), the Bethe eigenstates of the $t\!-\!0$~Hamiltonian~(\ref{t_0_Hamiltonian}) are given by the product of the eigenstate of $N = N_{\uparrow} + N_{\downarrow}$ non-interacting spinless fermions on $L$ sites and the eigenstate of the XX spin chain in an auxiliary lattice of~$M = N_{\downarrow}$ sites~\cite{Abarenkova1997, Izergin1998b}.  The size of this sector of the Hilbert space is~$\binom{L}{N} \times \binom{N}{M}$, where $\binom{X}{Y}$ is the binomial coefficient. The conservation of the spin pattern further reduces the dimensionality of the sector to
\be \label{Hilber_space_sector_dim_calD}
	{\cal D} = {\mathcal Z}_{ {\boldsymbol \sigma} }  \binom{L}{N},
\ee
where ${\mathcal Z}_{ {\boldsymbol \sigma} }$ is the length of the orbit of a spin pattern~${\boldsymbol \sigma} = \{ \sigma_1, \ldots, \sigma_N \}$ under the action of the cyclic group~$C_N$.
The value of~${\mathcal Z}_{ {\boldsymbol \sigma} }$ for a given spin pattern ${\boldsymbol \sigma}$ can be immediately read off from the {\it periodicity} of the pattern. Therefore, for $M>0$ one has $2 \leq {\cal Z}_{{\bs \sigma}} \leq N$, and the lower bound is achieved for the N\'eel state, whereas the upper bound is realized for a generic aperiodic spin pattern. Obviously, in the spin-polarized case $(M=0$ or $M=N)$ one has~${\mathcal Z}_{ {\boldsymbol \sigma} } = 1$. In what follows we always work in the sector with the fixed spin pattern.

\subsection{Eigenstates of the $t-0$ model} \label{eigenstates_t0_fixed_pattern}

In the Hilbert space sector with the fixed spin pattern~${\boldsymbol \sigma} = \{ \sigma_1, \ldots, \sigma_N \}$, which consists of~$N$ spins in total and contains~$M$ down spins, the eigenstate of the $t\!-\!0$ Hamiltonian can be written as~\cite{Mielke1991}
\begin{multline} \label{t_0_eigenstates}
	\ket{\Psi_{t-0}^{ {\boldsymbol \sigma} }( {\boldsymbol k}, \varphi )} = \sum_{ 1 \leq x_1 <  \ldots < x_N \leq L }  \psi_{ {\boldsymbol k},\varphi }({\boldsymbol x})   \\
	\times \sum_{j=0}^{ {\cal Z}_{ {\boldsymbol \sigma}} -1 } \chi_{\varphi}^{ {\boldsymbol \sigma} }(j) \prod_{l=1}^N c_{x_l, \sigma_{l+j}}^{\dag} \ket{0},
\end{multline}
where~$\psi_{ {\boldsymbol k},\varphi }({\boldsymbol x})$ and $\chi_{\varphi}^{ {\boldsymbol \sigma} }(j)$ are the wave functions of the charge and spin degrees of freedom, respectively, and the spin indices satisfy $\sigma_{j+ {\cal Z}_{ {\boldsymbol \sigma}} } \equiv \sigma_j$. 
The eigenstates~(\ref{t_0_eigenstates}) are parametrized by the set~${\bs k} = \{ k_1, \ldots, k_N \}$ of~$N$ momenta of the charge degrees of freedom and the {\it total} spin quasi-momentum~$\varphi$.
The charge part of the many-body wave function corresponds to the Slater determinant of~$N$ free spinless fermions on a ring penetrated by the magnetic flux~$\varphi$ and reads
\be \label{charge_wf_t0}
	\psi_{ {\boldsymbol k},\varphi }({\boldsymbol x}) = \det_{1 \leq a, b \leq N} \Bigl\{\frac{1}{\sqrt{L}}  \exp\bigl[ i (k_a + \varphi/L ) x_b \bigr] \Bigr\}.
\ee
On the other hand, the spin part of the many-body wave function corresponds to a {\it single} free particle with the quasi-momentum~$- \varphi$:
\be \label{spin_wf_t0} 
	\chi_{\varphi}^{ {\boldsymbol \sigma} }(j) = \frac{1}{ \sqrt{ {\cal Z}_{ {\boldsymbol \sigma} } }  }\exp\{ - i \varphi j \}.
\ee
Due to the periodic boundary conditions the charge rapidities~$k_a$ and the spin quasi-momentum~$\varphi$ satisfy the quanization conditions~$\exp\{ i k_a L \} = 1$, with $1\leq a \leq N$, and~$\exp\{- i \varphi {\cal Z}_{\boldsymbol \sigma} \} = 1$, so that we have
\begin{subequations} \label{t0_quantization}
\begin{align}
	k_a & = \frac{2\pi}{L} \kappa_a, 	&& \kappa_a \in\{ 0,1, \ldots, L-1\}, \\
    \varphi &= \frac{2\pi}{ {\cal Z}_{{\bs \sigma}} } s, &&  0 \leq s \leq  {\cal Z}_{{\bs \sigma}} -1.
\end{align}
\end{subequations}
The energy of the eigenstate~(\ref{t_0_eigenstates}) is given by
\be \label{t0_spectrum}
	E({\boldsymbol k}, \varphi)  = -2 t \sum_{a = 1}^N \cos \left(k_a + \varphi/ L \right).
\ee
The eigenstates~(\ref{t_0_eigenstates}) are orthonormal and form a complete set in the Hilbert space sector with the fixed spin pattern. 

We note that the form~(\ref{t_0_eigenstates}) of the eigenstates in the sector with fixed spin pattern~${\bs \sigma}$ is less known as compared to the eigenstates belonging to the sector with the fixed values of $N$ and $M$. For the sake of completeness we discuss the relation between the two sectors in Appendix~\ref{A:t0_BA_wf_NM_sec}. Here we only would like to emphasize that the advantage of the sector with fixed ${\bs \sigma}$ is that {\it both} the spectrum~(\ref{t0_spectrum}) {\it and} the eigenstates~(\ref{t_0_eigenstates}) depend on spin degrees of freedom via a single parameter -- {\it total} quasimomentum $\varphi$. This is no longer true for the eigenstates in the sector with fixed $N$ and $M$, in which case only the spectrum depends solely on $\varphi$, whereas the eigenstates depend on $M$ distinct spin quasimomenta. 


\section{Mapping of the $t-0$ model onto free fermions} \label{S:t0_model_mapping}

Remarkably, the infinite-$U$ Hubbard ($t\!-\!0$) model admits an elementary solution that does not require the use of Bethe ansatz. This solution relies on a certain decomposition of the spin-$1/2$ fermionic operators~$c_{j, \sigma}$ into the charge and spin degrees of freedom, and a unitary transformation that eliminates the spin dependence of the~$t\!-\!0$ Hamiltonian~(\ref{t_0_Hamiltonian}). Below we outline the main steps of this alternative approach.

\subsection{Fermionic operator mapping}

At the first step, one represents the operators~$c_{j,\sigma}$, $c_{j,\sigma}^{\dag}$ of physical {\it spinful} fermions in terms of the operators of {\it spinless} fermions~$f_j$, $f_j^{\dag}$ and the Pauli matrices~$\sigma_j^{\alpha}$, representing the charge and spin variables, respectively. The operators~$f_j$ and $f_j^{\dag}$ satisfy the canonical anticommutation relations~$\{ f_j, f_k^{\dag} \} = \delta_{j,k}$ and commute with the Pauli matrices,~$[f_{j}, \sigma_{k}^{\alpha}] = [f_{j}^{\dag}, \sigma_{k}^{\alpha}] = 0$. Equivalent forms of this mapping were independently obtained in a large number of works (see e.g.~\cite{Khaliullin1990, Richard1993, Wang1994, Chandler1997, Ostlund2006, Kumar2008}), 
and one has the following operator mapping
\be \label{c_to_f_sigma} 
\begin{split}
    c_{j, \uparrow}^{\dag} &= \left( f_j^{\dag} + (-1)^{j+1} f_j \right)  \sigma_j^{+} , \\
    c_{j, \downarrow}^{\dag} &= f_j^{\dag} \frac{1-\sigma_j^z}{2} + (-1)^j f_j \frac{1+\sigma_j^z}{2},
\end{split}
\ee
where~$\sigma_{j}^{+} = ( \sigma_j^x + i \sigma_j^y )/2$. 
Let us emphasize that the mapping in Eq.~(\ref{c_to_f_sigma}) yields a canonical transformation of the spin-$1/2$ fermionic operators~$c_{j,\sigma}$ and $c_{j,\sigma}^{\dag}$~\cite{mapping_note}.
From Eq.~(\ref{c_to_f_sigma})  one immediately obtains that the number operators of spinful fermions become
\be \label{n_updown_to_sigma_f}
	n_{j,\uparrow} = \frac{1}{2}\left( 1+ \sigma_{j}^{z} \right), \qquad n_{j, \downarrow} = n_{j,\uparrow} - {\cal N}_j \sigma_{j}^z,
\ee
where we introduced the number operator of spinless fermions
\be \label{N_spinless_fermions}
	{\cal N}_j = f_j^{\dag} f_j.
\ee 
In order to complete the mapping one has to specify the transformation of the local Hilbert space in Eq.~(\ref{4d_local_Hilbert_space}). 
Following Refs.~\cite{Nocera2008, Tartaglia2021}, for the physical vacuum~$\ket{0}_j$ we take
\be \label{vac_mapping}
	\ket{0}_j = \ket{ \fullmoon }_j \ket{ \Downarrow }_j,
\ee
where $\ket{\fullmoon}_j$ is the vacuum of spinless fermions,  and $\ket{\Downarrow}_j$ is the spin-down state.
Using Eq.~(\ref{c_to_f_sigma}), for the states with one fermion we obtain
\be
\begin{split} \label{local_states_one_fermion_at_most}
	c_{j, \downarrow}^{\dag} \ket{0}_j  &=  \phantom{\sigma_j^{+}} f_j^{\dag} \ket{ \fullmoon }_j \ket{ \Downarrow }_j = \ket{\newmoon}_j \ket{ \Downarrow }_j, \\
	c_{j, \uparrow}^{\dag} \ket{0}_j  &=  f_j^{\dag} \sigma_j^{+}  \ket{ \fullmoon }_j \ket{ \Downarrow }_j = \ket{\newmoon}_j \ket{ \Uparrow }_j,
\end{split}
\ee
where $\ket{\newmoon}_j = f_j^{\dag} \ket{\fullmoon}$ and $\ket{\Uparrow}_j = \sigma_j^+ \ket{\Downarrow}_j$.
Finally, for the doubly occupied site $c_{j,\uparrow}^{\dag}c_{j,\downarrow}^{\dag}\ket{0}_j$ we use Eq.~(\ref{c_to_f_sigma}) and write $c_{j, \uparrow}^{\dag} c_{j, \downarrow}^{\dag} = (-1)^{j+1} \left( 1 - {\cal N}_j \right) \sigma_j^{+}$. This immediately yields the mapping
\be
    c_{j,\uparrow}^{\dag}c_{j,\downarrow}^{\dag}\ket{0}_j = (-1)^{j+1} \ket{ \fullmoon }_j \ket{ \Uparrow }_j.
\ee


\subsection{Mapping of the $t\!-\!0$ model and elimination of the spin degrees of freedom}

Applying the transformation in Eq.~(\ref{c_to_f_sigma})  to the $t\!-\!0$ Hamiltonian~(\ref{t_0_Hamiltonian_v2}), one obtains~\cite{Kumar2009}
\be \label{t_0_H_to_sigma_f}
	H_{t-0} = - t \, \sum_{j=1}^{L} P_{j,j+1} \left( f_j^{\dag} f_{j+1} + \text{H.c.} \right),
\ee
where the spin degrees of freedom enter the Hamiltonian~(\ref{t_0_H_to_sigma_f}) only via the permutation operator
\be \label{perm_op}
	P_{i,j} = \frac{1}{2} \left( 1 + {\boldsymbol \sigma}_{i} \cdot {\boldsymbol \sigma}_{j} \right),
\ee
with~${\boldsymbol \sigma}_j = \{ \sigma_j^x, \sigma_j^y, \sigma_j^z \}$ being the vector of Pauli matrices.

Quite remarkably, the spin degrees of freedom can be effectively eliminated from the Hamiltonian~(\ref{t_0_H_to_sigma_f}), as was first shown by Kumar for the  case of open boundary conditions~\cite{Kumar2008, Kumar2009, Kumar2013} and later extended to the case of periodic boundary conditions in Ref.~\cite{Brzezicki2014}. This is achieved with the help of the unitary transformation
\be \label{Kumar_unitary}
	{\cal U} = \prod_{j=2}^{L} U_{j}, \quad U_{j} = \left( 1 - {\cal N}_{j} \right) + {\cal N}_{j} \, T_{j},
\ee
where ${\cal N}_j$ is given by Eq.~(\ref{N_spinless_fermions}) and the unitary operator~$T_{j}$ acts on the spin degrees of freedom.  Explicitly it is given by
\be \label{T_j_shift_op}
	T_{j} = P_{j, j-1} P_{j-1, j-2} \ldots P_{2,1}, \quad 2 \leq j \leq L.
\ee
The operator~$T_m$ satisfies the relations
$
	T_m^m =  1, \quad T_m^{\dag} = T_m^{-1}= T_m^{m-1}, 
$
and acts on the spin operators as
\be \label{T_j_action}
\begin{aligned}
	&T_{m}^{\dag} \, {\boldsymbol \sigma}_k \, T_{m} = \begin{cases}
		 {\boldsymbol \sigma}_{k+1}, &  k < m , \\
		 {\boldsymbol \sigma}_{k},  & k > m, \\
		 {\boldsymbol \sigma}_{1}, & k = m.  \\
		 \end{cases}
\end{aligned}
\ee
Since the operator~$T_m$ is unitary, we can write
\be \label{shift_op_exp_momentum}
	T_m = e^{i \Pi_m}, \qquad \Pi_m = \Pi_m^{\dag},
\ee
which for $m=L$ becomes the operator of lattice momentum~$\Pi_L$~\cite{Hubbard_book}. 

The unitary transformation with the operator~${\cal U}$ in Eq.~(\ref{Kumar_unitary}) reduces the Hamiltonian~(\ref{t_0_H_to_sigma_f}) to 
\be \label{H_tb}
	 H_{\text{tb}} = {\cal U}^{\dag}\, H_{t-0} \, {\cal U} =  -t \sum_{j=1}^{L-1} \left( f_{j}^{\dag} f_{j+1} + \text{H.c.} \right) + H_{B},
\ee
where the term~$H_{B}$ is given by
\be \label{boundary_link_hopping_initial_form}
	H_{B} = - t \, {\cal U}^{\dag} P_{L,1} \left(f_{L}^{\dag} f_1 + f_1^{\dag} f_L \right) {\cal U},
\ee
and it vanishes for open boundary conditions.
In the sector with the fixed number of particles~$N$, the expression in~Eq.~(\ref{boundary_link_hopping_initial_form}) can be significantly simplified (see Appendix~\ref{A:boundary_link_hopping} for the derivation), and one has
\be \label{boundary_link_hopping_v2}
	H_{B} = -t \left( e^{ - i \tilde \Pi_{L - N} } e^{ - i \, \Pi_N}  f_L^{\dag} f_1 + \text{H.c.} \right),
\ee
where we denoted $\tilde \Pi_{L - N} =  T_L^{-N} \Pi_{L - N} T_L^{N}$ and for $N=L-1$ it is understood $\Pi_1 \equiv 0$.
Then, taking into account the translational invariance and replacing in Eq.~(\ref{boundary_link_hopping_v2}) the operators $\Pi_p$ with their eigenvalues ${\cal K}_p = 2\pi k/p$, where $k \in \{ 0, \ldots, p-1 \}$, one immediately reduces Eq.~(\ref{t_0_H_to_sigma_f}) to
\begin{multline} \label{H_tb_quasi_pbc}
	H_{\text{tb}} = -t \sum_{j=1}^{L-1} \left( f_{j}^{\dag} f_{j+1} + \text{H.c.} \right) \\
  - t \left( e^{- i \left( {\cal K}_N + {\cal K}_{L-N} \right)} f_L^{\dag} f_1 + \text{H.c.} \right).
\end{multline}
Thus, we see that the Hamiltonian~(\ref{H_tb_quasi_pbc}) describes free spinless fermions in a ring with twisted (quasiperiodic) boundary conditions. The twisting phase~${\cal K}_N + {\cal K}_{L-N}$ originates from the coupling between the charge and spin degrees of freedom. In the next subsection we show that this phase is nothing else than the spin quasimomentum~$-\varphi$ entering the Bethe Ansatz eigenstate~(\ref{t_0_eigenstates}) of the~$t\!-\!0$ model. 

\subsection{Eigenstates of $H_{\text{tb}}$}

One can now easily check that the eigenstates of the tight-binding Hamiltonian $H_{\text{tb}}$ in Eq.~(\ref{H_tb_quasi_pbc}) are
\begin{multline} \label{Psi_tb_gen}
	\ket{\Psi_{\text{tb}}^{ {\boldsymbol \sigma} } \left( {\boldsymbol k}, \varphi \right) }  = {\cal U}^{\dag} \ket{ \Psi_{t-0}^{{\boldsymbol \sigma}} \left( {\boldsymbol k}, \varphi \right) }
	 = \ket{ \psi_{ {\boldsymbol k}, \varphi } } \otimes \ket{ \chi_{\varphi}^{ {\boldsymbol \sigma} } },
\end{multline}
where the eigenstate of the $t\!-\!0$ model $\ket{ \Psi_{t-0}^{{\boldsymbol \sigma}} \left( {\boldsymbol k}, \varphi \right) }$ is given by Eq.~(\ref{t_0_eigenstates}). The states~$\ket{ \psi_{ {\boldsymbol k}, \varphi } }$ and~ $\ket{ \chi_{\varphi}^{ {\boldsymbol \sigma} } }$ read
\begin{subequations} \label{Psi_tb_spin_charge} 
\begin{align}
	\ket{ \psi_{ {\boldsymbol k}, \varphi } } &= \sum_{ 1\leq x_1 < \ldots < x_N \leq L } \psi_{ {\boldsymbol k}, \varphi } ( {\boldsymbol x} ) \ket{ {\bs x} } , \\
	\ket{ \chi_{\varphi}^{ {\boldsymbol \sigma} } } & = \sum_{j=0}^{{\cal Z}_{ {\boldsymbol \sigma} } - 1} \chi_{\varphi}^{ {\boldsymbol \sigma} }(j) \ket{ {\cal C}^j {\bs \sigma} }, \label{Psi_tb_spin} 
\end{align}
\end{subequations}
where the wave functions~$\psi_{ {\boldsymbol k}, \varphi } ( {\boldsymbol x} )$ and~$\chi_{\varphi}^{ {\boldsymbol \sigma} }(j)$ given by~Eqs.~(\ref{charge_wf_t0}) and~(\ref{spin_wf_t0}), respectively. 
The states~$\ket{{\bs x}}$ and $\ket{ {\cal C}^j {\bs \sigma} }$ in Eq.~(\ref{Psi_tb_spin_charge}) read
\begin{subequations} 
\begin{align}
    \ket{ {\bs x} } &= \prod_{l=1}^N f_{x_l}^{\dag} \ket{\fullmoon}, \label{generic_charge_eigenstates_product_state} \\
    \ket{ {\cal C}^j {\bs \sigma} } &=  \bigl| \underbrace{\sigma_{N+j}, \ldots, \sigma_{1+j} }_{N}, \underbrace{ \Downarrow, \ldots \Downarrow}_{L-N} \bigl\rangle, \label{generic_spin_product_state}
\end{align}
\end{subequations}
and in Eqs.~(\ref{Psi_tb_gen}) and~(\ref{Psi_tb_spin_charge}) we took into account that for a product state~$\prod_{l =1}^N c_{x_l, \sigma_{l+j}}^{\dag} \ket{0}$ one has~\cite{Tartaglia2021}
\be \label{x_sigma_ket}
	{\cal U}^{\dag} \prod_{l =1}^N c_{x_l, \sigma_{l+j}}^{\dag} \ket{0} \equiv \ket{ {\boldsymbol x} }\!\otimes\! \ket{ {\cal C}^j {\boldsymbol \sigma}}.
\ee
We note that the following relations hold for the eigenstates of the Hamiltonian~(\ref{H_tb_quasi_pbc}):
\be
\begin{split}
	&e^{- i \Pi_N} \ket{\Psi_{\text{tb}}^{ {\boldsymbol \sigma} } \left( {\boldsymbol k}, \varphi \right) }  = e^{i \varphi} \ket{\Psi_{\text{tb}}^{ {\boldsymbol \sigma} } \left( {\boldsymbol k}, \varphi \right)} , \\
	&e^{ - i \, \tilde \Pi_{L - N} } \ket{ \Psi_{\text{tb}}^{ {\boldsymbol \sigma} } \left( {\boldsymbol k}, \varphi \right) }  = \ket{\Psi_{\text{tb}}^{ {\boldsymbol \sigma} } \left( {\boldsymbol k}, \varphi \right)},
\end{split}
\ee
where~$T_m$ is given by Eq.~(\ref{T_j_shift_op}). Therefore, for the twisting phase in Eq.~(\ref{H_tb_quasi_pbc}) we have
\be \label{twisting_phase}
	- \left( {\cal K}_N + {\cal K}_{L-N} \right) = \varphi,
\ee
where $\varphi$ is the spin quasimomentum [see Eqs.~(\ref{spin_wf_t0}) and~(\ref{t0_quantization})]. In other words, the spin quasimomentum plays the role of an effective magnetic flux for the charge degrees of freedom.


\section{Disordered $t\!-\!0$ model} \label{S:Hubbard_disorder}

We now turn to the one-dimensional $t\!-\!0$ model in the present of disorder and consider the operators
\begin{subequations} \label{H_D_gen}
\begin{align}
	V_{\text{charge}} &= \sum_{j=1}^L \varepsilon_j \left( n_{j,\uparrow} + n_{j,\downarrow} \right), \label{random_potential_def} \\
 	V_{\text{spin}} &= \sum_{j=1}^L h_j \left( n_{j,\uparrow} - n_{j,\downarrow} \right), \label{random_magnetic_field_def}
\end{align}
\end{subequations}
with~$\varepsilon_j$ and $h_j$ being the random on-site potential and magnetic field, respectively. For our purposes the specific forms of the distributions for $\varepsilon_j$ and $h_j$ are not important and can be arbitrary. 
Projecting the operators $V_{\text{charge}}$ and $V_{\text{spin}}$ onto the subspace with no doubly occupied sites, we take into account that one has ${\mathsf P} V_{\text{charge(spin)}} {\mathsf P} = V_{\text{charge(spin)}} {\mathsf P}$, where ${\mathsf P}$ is given by Eq.~(\ref{strong_coupling_P}). Therefore, one can safely omit the projector~${\mathsf P}$ since all operators act on the projected Hilbert space. We then add both terms in Eq.~(\ref{H_D_gen}) to the $t\!-\!0$~Hamiltonian~(\ref{t_0_Hamiltonian}) and obtain
\be \label{H_t0_H_D}
	H_{\text{tot}} = H_{t-0} + V_{\text{charge}} + V_{\text{spin}},
\ee
Disorder of any type breaks translational invariance, and in addition the random magnetic field breaks the~$SU(2)$ symmetry. However, both the purely potential disorder~$V_{\text{charge}}$ and the random magnetic field~$V_{\text{spin}}$ conserve the numbers of spin-up and spin-down particles. Importantly, both types of disorder do {\it not} violate the conservation of the spin pattern. 

\subsection{Random potential} \label{S:potential_disorder}

We first consider the case of a purely potential (spin-independent) disorder, i.e. we put $V_{\text{spin}} \equiv 0$ in Eq.~(\ref{H_t0_H_D}).
Using the mapping in Eq.~(\ref{n_updown_to_sigma_f}) in order to express the number operators of spinful fermions $n_{j,\sigma}$ in terms of the charge and spin degrees of freedom, one finds
\be \label{pot_disorder_2}
	V_{\text{charge}} = \sum_{j=1}^L \varepsilon_j \left( 1 +  \left( 1 - {\cal N}_j \right) \sigma_j^z \right).
\ee
Then, we take into account that~$ {\cal N}_j \ket{\newmoon}_j = \ket{\newmoon}_j$ and ${\cal N}_j \ket{\fullmoon}_j = 0$. Since in the limit of~$U \to +\infty$ the local Hilbert space is spanned by the three states given by Eqs.~(\ref{vac_mapping}) and~(\ref{local_states_one_fermion_at_most}), we see that the action of the operator~$(1 - {\cal N}_j) \sigma_j^z$ differs from zero only for the state~$\ket{\fullmoon}_j \ket{\Downarrow}_j$ and yields~$(-1)\ket{\fullmoon}_j \ket{\Downarrow}_j$. Therefore, in Eq.~(\ref{pot_disorder_2}) one can simply replace~$\sigma_j^z \to -1$, 
and we obtain
\be \label{potential_disorder_f_sigma}
	V_{\text{charge}} = \sum_{j=1}^L \varepsilon_j {\cal N}_j.
\ee

One then needs to perform a unitary transformation of~$V_{\text{charge}}$ in Eq.~(\ref{potential_disorder_f_sigma}) with the operator~${\cal U}$ from Eq.~(\ref{Kumar_unitary}).  Taking into account that~${\cal N}_j$ and ${\cal U}$ commute, one has~${\cal U}^{\dag} V_{\text{charge}} {\cal U} = V_{\text{charge}}$, and the total Hamiltonian becomes
\begin{equation} \label{H_potential_disorder_final}
	H_{\text{A}} = {\cal U}^{\dag} H_{\text{tot}} {\cal U} = H_{\text{tb}} + V_{\text{charge}},
\end{equation}
where $H_{\text{tb}}$ is given by Eqs.~(\ref{H_tb_quasi_pbc}) and (\ref{twisting_phase}). The Hamiltonian~(\ref{H_potential_disorder_final}) describes one-dimensional non-interacting spinless fermioins with quasi-periodic boundary conditions subject to the potential disorder. It is well known that this model exhibits Anderson (single-particle) localization at an arbitrarily weak disorder strength. However, since the disorder potential in Eq.~(\ref{H_potential_disorder_final}) only acts on the charge degrees of freedom, any cyclic permutation of the spin pattern has the same energy, just like in the disorder-free case. Therefore, the spin eigenstates remain plane waves, as given by Eqs.~(\ref{spin_wf_t0}) and~(\ref{Psi_tb_spin_charge}). Thus, the eigenstates of the Hamiltonian~(\ref{H_potential_disorder_final}) are labeled by the set of coordinates~${\boldsymbol r} = \{r_1, \ldots, r_N \}$ of {\it localized} single-particle charge degrees of freedom and the total quasimomentum~$\varphi$ of the spin degrees of freedom, so that one has
\be \label{Anderson_factorized_eigenstate}
	\ket{\Psi^{ {\boldsymbol \sigma} } ( {\boldsymbol r}, \varphi )} = \ket{\psi_{\varphi}({\boldsymbol r}) } \otimes \ket{ \chi_{\varphi}^{ {\boldsymbol \sigma} }  },
\ee
where the spin part~$\ket{ \chi_{\varphi}^{ {\boldsymbol \sigma} }  }$ is given by Eq.~(\ref{Psi_tb_spin}) and the charge part reads
\be
	\ket{\psi_{\varphi} ( {\bs r} ) } = \frac{1}{N!} \sum_{x_1 = 1}^L \ldots \sum_{x_N = 1}^L \psi_{ \varphi }( {\bs r}, {\bs x} ) \ket{ {\bs x} },
\ee
with the wave function $\psi_{ \varphi }( {\bs r}, {\bs x} )$ being the Slater determinant of spinless fermions
\be \label{Anderson_wf_Slater_det}
	\psi_{ \varphi }( {\bs r}, {\bs x} ) =  \det_{1\leq a, b\leq N}\left\{ \psi_{\varphi}( r_a, x_b ) \right\},
\ee
where $\psi_{\varphi}( r_a, x)$ is a wave function of a single particle localized around the site~$r_a \in \{ x_1, \ldots, x_N \}$ of the chain.

\subsection{Random magnetic field} \label{S:random_field}
 
Let us now turn to the case of non-zero random magnetic field.
Under the mapping~(\ref{n_updown_to_sigma_f}), the operator~$V_{\text{spin}}$ in Eq.~(\ref{H_D_gen}) becomes 
\be \label{random_field_f_sigma}
	V_{\text{spin}} = \sum_{j=1}^L h_j {\cal N}_j \sigma_j^z.
\ee
Thus, one can not eliminate the~$\sigma_j^z$ operator as was done in the case of potential disorder, since the operator~${\cal N}_j \sigma_j^z$ acts {\it differently} on the states $\ket{\newmoon}_j\ket{\Downarrow}_j$ and $\ket{\newmoon}_j\ket{\Uparrow}_j$. 
Then, performing the unitary transformation of~$V_{\text{spin}}$ with the operator~${\cal U}$ from Eq.~(\ref{Kumar_unitary}), we arrive at
\be \label{random_field_transformed_H}
	\tilde V_{\text{spin}} = {\cal U}^{\dag} V_{\text{spin}} {\cal U}  =  \sum_{j=1}^L h_j {\cal N}_j  \sigma_{N + 1-\nu_{1,j}}^z,	
\ee
where we used the results of Appendix~\ref{A:sigma_transform} for ${\cal U}^{\dag} \sigma_j^z {\cal U}$ and introduced the counting function
\be \label{counting_function_definition}
	\nu_{1,j} = \sum_{k=1}^j {\cal N}_k,
\ee 
which counts the number of occupied sites in an interval of length $j$~\cite{Zvonarev_private}. A few comments regarding~$\nu_{1,j}$ are in order. First, the counting function obviously satisfies $\nu_{1,L} = N$. Second, due to the periodic boundary conditions, one has $\nu_{1+L, j+L} = \nu_{1, j}$. Finally, an extra care is needed with the first-quantized representation of the counting function. Consider the action of $\nu_{1,x_m}$ on the state $\ket{{\bs x}} = f_{x_1}^{\dag} \ldots f_{x_m}^{\dag} \ldots f_{x_N}^{\dag} \ket{\fullmoon}$. Clearly, we have
\be \label{counting_function_action} 
    \nu_{1, x_m} \ket{{\bs x}} = \sum_{k=1}^{x_m} \sum_{l=1}^N \delta_{k, x_l} \ket{{\bs x}}.
\ee
Therefore, for the set ${\bs x} = \{ x_1, \ldots, x_m, \ldots, x_N \}$ that can be ordered by some permutation $R \in {\cal S}_N$ as 
\be
    1 \leq x_{R1} < \ldots < x_{Rm} < \ldots < x_{RN} \leq L,
\ee
from Eq.~(\ref{counting_function_action}) one obtains~\cite{counting_function_example}
\be
    \nu_{1,x_m} \ket{{\bs x}} = Rm \ket{\bs x}.
\ee    
Thus, for an ordered sector $1 \leq x_1 < \ldots < x_l \ldots < x_N$, i.e. for $R = 1$, one simply has $\nu_{1, x_m} \ket{ {\bs x}} = m \ket{{\bs x}}$. This fact greatly simplifies treatment of the random magnetic field~$\tilde V_{\text{spin}}$.

From Eq.~(\ref{random_field_transformed_H}) we see that the random magnetic field results in a complicated nonlocal coupling between the charge and spin sector, so that for $V_{\text{spin}} \neq  0$ the Hamiltonian~(\ref{H_t0_H_D}) can not be factorized into the spin and charge parts. 

\section{Matrix elements of the random magnetic field}

In this section we proceed with deriving analytic expressions for matrix elements of the random magnetic field $\tilde V_{\text{spin}}$ in~Eq.~(\ref{random_field_transformed_H}), calculated between the eigenstates~(\ref{Anderson_factorized_eigenstate}) of 
the Anderson Hamiltonian~(\ref{H_potential_disorder_final}).

\subsection{General case}

First we observe that the random magnetic field~$\tilde V_{\text{spin}}$ in Eq.~(\ref{random_field_transformed_H}) does not couple sectors with different spin patterns~${\bs \sigma}$.
Then, using Eq.~(\ref{Anderson_factorized_eigenstate}), we have
\begin{multline} \label{random_field_matrix_element_1}
	\langle \Psi^{{\bs \sigma}} ( {\bs r}, \varphi ) | \tilde V_{\text{spin}} | \Psi^{{\bs \sigma}} ( {\bs r}^{\prime}, \varphi^{\prime} ) \rangle = \sum_{ {\bs x}, {\bs x}^{\prime} } 
    \sum_{j, j^{\prime}} \chi_{\varphi}^{{\bs\sigma}}{}^* (j) \chi_{\varphi^{\prime}}^{{\bs\sigma}}(j^{\prime}) \\
	\times  \psi_{\varphi}^*( {\bs r}, {\bs x} ) \psi_{\varphi^{\prime}}( {\bs r}^{\prime}, {\bs x}^{\prime})
	\; \langle {\bs x} ; {\mathcal C}^{j}{\bs \sigma} | \tilde V_{\text{spin}} | {\bs x}^{\prime} ;  {\mathcal C}^{j^{\prime}} {\bs \sigma} \rangle,
\end{multline}
where for brevity we denoted $| {\bs x} ; {\cal C}^j {\bs \sigma} \rangle \equiv \ket{{\bs x}} \otimes | {\cal C}^j {\bs \sigma}\rangle$.
Taking into account that the random magnetic field is given by~$\tilde V_{\text{spin}} = \sum_{j=1}^L h_j {\cal N}_j \sigma^z_{N-\nu_{1,j} + 1}$ [see Eq.~(\ref{random_field_transformed_H})], we obtain
\be \label{random_field_product_state_action}
	\tilde V_{\text{spin}} | {\bs x} ;  {\mathcal C}^j {\bs \sigma} \rangle 
 = \sum_{\ell = 1}^N h_{x_{\ell}} \epsilon\left( \sigma_{j + \nu_{1, x_{\ell}} } \right) \ket{ {\bs x}  ; {\mathcal C}^j {\bs \sigma} },
\ee
where $\epsilon(\Uparrow) = 1$,  $\epsilon(\Downarrow) = -1$, and we took into account that~$\sigma^z_{N - m + 1} \ket{{\cal C}^j{\boldsymbol\sigma}} = \epsilon( \sigma_{m+j} ) \ket{{\cal C}^j {\boldsymbol\sigma}}$, as follows from the definition of $\ket{{\cal C}^j {\boldsymbol\sigma}}$ in Eq.~(\ref{generic_spin_product_state}). Note that one can write
\be \label{epsilon_l_y_def}
	\epsilon( \sigma_{m} ) = 1 - 2 \sum_{k=1}^M \delta_{m, y_k},
\ee
where~$1\leq y_k \leq N$ are the coordinates of $M$ down spins in the spin pattern~${\boldsymbol \sigma} = \{ \sigma_1, \ldots, \sigma_N \} $.
Therefore, taking into account that the states~$\ket{ {\boldsymbol x} ; {\cal C}^j {\boldsymbol \sigma}}$ are orthonormal, 
we can write the matrix element~(\ref{random_field_matrix_element_1}) as
\begin{widetext}
\be \label{random_field_matrix_element_2}
\begin{split} 
	\langle \Psi^{{\bs \sigma}} ( {\bs r}, \varphi ) | \tilde V_{\text{spin}} | \Psi^{{\bs \sigma} } ( {\bs r}^{\prime}, \varphi^{\prime} ) \rangle 
	 &= 
	\frac{1}{N!} \sum_{x_1 = 1}^L\!\ldots\!\sum_{x_N = 1}^L 
		\sum_{\ell = 1}^N h_{x_{\ell}}  \psi_{\varphi}^*( {\bs r}, {\bs x} ) \psi_{\varphi^{\prime}}( {\bs r}^{\prime}, {\bs x}) 
		\times \sum_{j = 1}^{ {\cal Z}_{ {\bs \sigma} } } \epsilon\left( \sigma_{ j + \nu_{1,x_{\ell}}} \right) 
		\chi_{ \varphi }^{ {\bs\sigma} }{}^* (j) \chi_{\varphi^{\prime}}^{{\bs\sigma}}(j) \\
   &= \sum_{1 \leq x_1 < \ldots < x_N \leq L } 
		\sum_{\ell = 1}^N h_{x_{\ell}}  \psi_{\varphi}^*( {\bs r}, {\bs x} ) \psi_{\varphi^{\prime}}( {\bs r}^{\prime}, {\bs x}) 
		\times 
        \frac{ 1 }{ {\cal Z}_{ {\boldsymbol \sigma} } }  \sum_{ j = 1  }^{ {\cal Z}_{ {\boldsymbol \sigma} } } \epsilon\left( \sigma_{j + \ell } \right) e^{i (\varphi - \varphi^{\prime}) j },
\end{split}
\ee
where in the second equality we replaced the repeated summation over $x_1, \ldots, x_N$ with the ordered one and took into account that in the ordered sector $x_1 < \ldots < x_N$ the counting function becomes simply $\nu_{1, x_{\ell}} = \ell$, as can be seen from its definition~(\ref{counting_function_definition}).
Then, using~Eq.~(\ref{epsilon_l_y_def}) for $\epsilon(\sigma_{m})$, we can rewrite in Eq.~(\ref{random_field_matrix_element_2}) the sum over $j$ as
\begin{equation}
    \frac{ 1 }{ {\cal Z}_{ {\boldsymbol \sigma} } }  \sum_{ j = 1  }^{ {\cal Z}_{ {\boldsymbol \sigma} } } \epsilon\left( \sigma_{j + \ell } \right) e^{i (\varphi - \varphi^{\prime}) j } \\
     =   \delta_{\varphi, \varphi^{\prime}}
     - \frac{ 2 }{ {\cal Z}_{ {\boldsymbol \sigma} } } \sum_{k=1}^{ {\cal Z}_{ {\boldsymbol \sigma} } \frac{M}{N} } e^{ i (\varphi - \varphi^{\prime}) ( y_k - \ell ) },
\end{equation}
where~$1 \leq y_k \leq {\cal Z}_{ {\boldsymbol \sigma} }$ are the positions of down spins in the pattern~${\boldsymbol \sigma}$, and the summation is only over the first~${\cal Z}_{ {\boldsymbol \sigma} } M/N$ positions of down spins (${\cal Z}_{ {\boldsymbol \sigma} } M/N$ is always an integer).  
Then, taking into account that $\delta_{\varphi, \varphi^{\prime}} e^{-  i (\varphi - \varphi^{\prime})\ell } = \delta_{\varphi, \varphi^{\prime}}$, we reduce the matrix element~(\ref{random_field_matrix_element_2}) to the following (almost) factorized form
\be \label{random_field_matrix_element_3}
    	\langle \Psi^{{\bs \sigma}} ( {\bs r}, \varphi ) | \tilde V_{\text{spin}} | \Psi^{{\bs \sigma} } ( {\bs r}^{\prime}, \varphi^{\prime} ) \rangle 
	= 
    F_{ \varphi , \varphi^{\prime}  } ( {\bs r}, {\bs r}^{\prime} ) 
    S^{ {\bs \sigma} }_{ \varphi - \varphi^{\prime} },
\ee
where we used the results of Appendix~\ref{Slater_dets_ME_sym_many_body} and denoted the spin part~$S^{ {\bs \sigma} }_{ \varphi - \varphi^{\prime} }$ and the charge part~$F_{ \varphi , \varphi^{\prime} } ({\bs r}, {\bs r}^{\prime} )$ by
\begin{subequations} \label{S_F_gen_def}
\begin{align}
    S^{ {\bs \sigma} }_{ \varphi - \varphi^{\prime} }  &= \delta_{\varphi- \varphi^{\prime},0} - \frac{ 2 }{ {\cal Z}_{ {\boldsymbol \sigma} } } \sum_{k=1}^{ {\cal Z}_{ {\boldsymbol \sigma} } \frac{M}{N} } e^{ i (\varphi - \varphi^{\prime}) y_k }, \label{S_gen_def} \\
    F_{ \varphi , \varphi^{\prime} } ({\bs r}, {\bs r}^{\prime} ) &= \sum_{\ell = 1}^N e^{-  i (\varphi - \varphi^{\prime}) \ell } \sum_{1 \leq x_1 < \ldots < x_N \leq L } 
		 h_{x_{\ell}} \; \psi_{\varphi}^*( {\bs r}, {\bs x} ) \psi_{\varphi^{\prime}}( {\bs r}^{\prime}, {\bs x}) \notag \\
  &= e^{-i \frac{N+1}{2} (\varphi - \varphi^{\prime}) } \sum_{l=1}^N \sum_{s=1}^N (-1)^{l+s}  \sum_{x = 1}^L  h_x \psi^*_{\varphi}(r_{l}, x) \psi_{\varphi^{\prime}}(r_{s}^{\prime}, x) D^{(l,s)}_{\varphi, \varphi^{\prime}}( x | {\bs r}, {\bs r}^{\prime} ),   \label{F_gen_def} 
\end{align}
\end{subequations}
with the function~$D^{(l,s)}_{\varphi, \varphi^{\prime}}( x | {\bs r}, {\bs r}^{\prime} )$ given by
\begin{equation} \label{D_determinant_def}
    D^{(l,s)}_{\varphi, \varphi^{\prime}}( x | {\bs r}, {\bs r}^{\prime} ) = \det_{\substack{1\leq a,b \leq N \\ (a\neq l, b \neq s)}} 
    \Biggl\{ \sum_{y=1}^L e^{-i \,\text{sgn}(x-y) (\varphi - \varphi^{\prime}) /2} 
     \psi^*_{\varphi}(r_{a}, y) \psi_{\varphi^{\prime}}(r_{b}^{\prime},y)
    \Biggr\}.
\end{equation}
\end{widetext}
Note that the charge part of the matrix element~$F_{ \varphi , \varphi^{\prime} } ({\bs r}, {\bs r}^{\prime} )$ still depends on the spin quasimomenta~$\varphi$ and $\varphi^{\prime}$ , so that the charge and spin degrees of freedom are not fully decoupled. One can easily check that for the homogeneous magnetic field, $h_j = h$, the matrix element~(\ref{random_field_matrix_element_3}) reduces to
$
    \langle \Psi^{{\bs \sigma}} ( {\bs r}, \varphi ) | \tilde V_{\text{spin}} | \Psi^{{\bs \sigma}} ( {\bs r}^{\prime}, \varphi^{\prime} ) \rangle 
	= (N-2M) \,h\, \delta_{\varphi,\varphi^{\prime}} \delta_{ {\bs r}, {\bs r}^{\prime}} 
$, as it should. In Appendix~\ref{app_ME_special_cases} we discuss some other special cases of the matrix elements~(\ref{random_field_matrix_element_3}).

The results presented in Eqs.~(\ref{random_field_matrix_element_3}) -- (\ref{D_determinant_def}) constitute one of the main findings of the present work and provide analytic expressions for the matrix elements of the random magnetic field (\ref{random_magnetic_field_def}) calculated between the Anderson localized (single-particle) eigenstates of the Hamiltonian~(\ref{H_potential_disorder_final}).
One can see that the random magnetic field strongly hybridizes the localized eigenstates~$\ket{ \Psi^{{\bs \sigma}}({\bs r, \varphi}) }$ (since there is an extensive number of non-zero off-diagonal matrix elements), which may lead to their delocalization. 
Let us now make a remark. Note that the matrix elements~(\ref{random_field_matrix_element_3})~--~(\ref{D_determinant_def}) depend on a {\it realization}~$\{ h_j \}_{j=1}^L$ of the random magnetic field. 
No assumptions were made on the disorder distribution so that our results are valid for any distribution of the random magnetic field. Thus, the next step is to investigate the statistical properties of the matrix elements for various random magnetic distributions, e.g. the uniform distribution. However, this is beyond the scope of the present paper, and we leave this for future work.

\section{Conclusions} \label{S:Conclusions}

In conclusion, we have analytically studied the one-dimensional Hubbard model for two-component fermions with infinitely strong on-site repulsion (the $t\!-\!0$ model) in the presence of disorder. 
We have used the fact that the spin and charge degrees of freedom of this model are effectively factorized. In the absence of disorder, this factorization allows one to effectively integrate out the spin degrees of freedom from the~$t\!-\!0$ model.  The resulting Hamiltonian becomes simply the tight binding model for free non-interacting spinless fermions representing the charge degrees of freedom. The spin degrees of freedom are also free, and for open boundary conditions they are completely eliminated~\cite{OBC_remark}. For periodic boundary conditions that we consider, the situation is more peculiar and the spin quasi-momentum twists the boundary conditions for spinless fermions.
We then investigated the influence of disorder on the effective model and demonstrated that the type of disorder drastically changes the nature of emerging phases. 
The case of potential (spin-independent) disorder can be treated as a single-particle problem which exhibits Anderson localization for the {\it charge} degrees of freedom. At the same time, the spin degrees of freedom are not affected by the potential disorder and they remain extended. 
We then considered the case of spin-dependent disorder (random magnetic field).
Unlike in the case of the potential disorder, the spin and charge degrees of freedom can no longer be factorized and one deals with a genuine many-body system with disorder. 
We have analytically calculated the matrix elements of the random magnetic field in the eigenbasis of the $t-0$ Hamiltonian with potential disorder, i.e. the Anderson-localized single-particle states.
The resulting matrix elements, given by Eqs.~(\ref{random_field_matrix_element_3})~--~(\ref{D_determinant_def}), are represented by determinants of a matrix whose entries are simple functions involving the localized single-particle states. Note that the matrix elements depend on a realization of the random magnetic field, so that our results are valid for any disorder distribution. It would be interesting to investigate the statistical properties of the matrix elements for various distributions of the random magnetic field. To do this one needs to relate the statistical properties of the localized single-particle states to those of the matrix elements of the random magnetic field. Note that the form of the matrix elements~(\ref{random_field_matrix_element_3})~--~(\ref{D_determinant_def}) is especially convenient for this task, since the dependence on single-particle states is rather transparent.

Our analysis of the matrix elements shows that the spin-dependent disorder strongly couples the charge and spin degrees of freedom and hybridizes the localized single particle states, which may lead to the many-body localization-delocalization  transition. 
This expectation is in agreement with the numerical observation of Ref~\cite{Bahovadinov2022} that the random magnetic field causes the localization-delocalization transition. In order to perform a direct comparison between our results and those of Ref~\cite{Bahovadinov2022}, one needs to investigate the statistical properties of the matrix elements with a uniformly distributed random magnetic field (as in Ref.~\cite{Bahovadinov2022}). Thus, the results of the present work are important for further analytical studies of many-body localization in the Hubbard model.
Our findings are also relevant for ongoing experiments with ultracold gases in optical lattices, where the type of disorder can be controlled.  

Finally, let us mention that it would be also interesting to investigate the $1/U$ corrections for the case of a very large but finite onsite repulsion between different spin components. However, in this case one has to deal with the significantly more complicated eigenstates of the finite-$U$ Hubbard model~\cite{Ogata1990, Hubbard_book}. We expect that in this case the many-body localization transition would be present already with a purely potential (spin-independent) disorder.

\acknowledgements

We thank O. Gamayun and M. Zvonarev for numerous useful discussions. This work was supported by the Russian Science Foundation Grant No. 20-42-05002 (Sec.~\ref{S:1D_Hubbard_overview} and~\ref{S:random_field}; the study of the effect of random magnetic field), the Russian Roadmap on Quantum Computing (Sec.~\ref{S:potential_disorder}; the study of potential disorder), Contract No. 868-1.3-15/15-2021, October 5, 2021); and by the Priority 2030 program at the National University of Science and Technology ``MISIS” under the project K1-2022-027 (calculation of matrix elements). MSB thanks Basic Research Program of HSE
for the provided support.

\appendix
\begin{widetext}

\section{Eigenstates of the $t-0$ model in the sector with fixed $N$ and $M$} \label{A:t0_BA_wf_NM_sec}

In this Appendix we discuss the relation between the eigenstates of the $t-0$ model in the sector with a fixed spin pattern~${\bs \sigma} = \{ \sigma_1, \ldots \sigma_N \}$ (see Sec.~\ref{eigenstates_t0_fixed_pattern}) and those in the sector with fixed total number of particles $N$ and number of spin-$\downarrow$ particles~$M$. Obviously, the latter is larger in size (except for the cases $M=0,1,N-1,N$, in which the sector with fixed ${\bs \sigma}$ and the one with fixed $N,M$ coincide), since one can construct various patterns~${\bs \sigma}$ for fixed $N$ and $M$.

In the sector with fixed values of~$N$ and~$M$ the Bethe Ansatz eigenstates of the $t-0$~Hamiltonian~(\ref{t_0_Hamiltonian}) can be written as~\cite{Hubbard_book, Abarenkova1997}
\be \label{t_0_eigenstates_fixed_NM_sector}
	\ket{\Psi_{t-0}^{N,M}( {\boldsymbol k}, {\boldsymbol \lambda} )} =\sum_{ 1 \leq x_1 < \ldots < x_N \leq L }  \; \sum_{\substack{ \sigma_1 , \ldots, \sigma_N = \uparrow, \downarrow \\ \sum_{s=1}^N \delta_{\sigma_s, \downarrow}= M } } \psi_{{\bs k}, \varphi}({\bs x}) \xi_{ {\bs \lambda} }( {\bs \sigma} )
	 c_{x_1, \sigma_1}^{\dag} \ldots c_{x_N, \sigma_N}^{\dag}  \ket{0},
\ee
where the summation over $\sigma_1, \ldots, \sigma_N$ includes only the terms containing exactly $M$ down spins, $\psi_{{\bs k}, \varphi}({\bs x})$ is the charge wave function given by Eq.~(\ref{charge_wf_t0}) in the main text,  and $\xi_{ {\bs \lambda} }( {\bs \sigma} )$ is the spin wave fucntion that reads
\be
    \xi_{ {\bs \lambda} }( {\bs \sigma} ) = \det_{1 \leq a, b \leq M} \left\{ \frac{1}{\sqrt{N}} e^{i \lambda_a y_b( {\bs \sigma} )} \right\},
\ee
where $y_m( {\boldsymbol \sigma}  )$ gives the position of the $m$th down spin in the sequence~${\boldsymbol \sigma} = \{ \sigma_1, \ldots, \sigma_N \}$.
The eigenstates~(\ref{t_0_eigenstates}) are parametrized by the sets of~$N$ charge rapidities~${\boldsymbol k} = \{ k_1, \ldots , k_N \}$ and $M$ spin rapidities~${\boldsymbol \lambda} = \{ \lambda_1, \ldots \lambda_M \}$, with the total spin quasimomentum being
\be \label{Total_spin_rapidity_def} 
	\varphi = \sum_{m=1}^M \lambda_m \text{ mod } 2\pi.
\ee
The charge and spin rapidities satisfy the quantization condition
\be
    e^{ik_a L} = 1 \quad (1 \leq a \leq N), \qquad \qquad e^{i \lambda_b N} = (-1)^{M+1} \quad (1 \leq b \leq M),
\ee
known as the Bethe Ansatz equations. Their solutions are given by
\be \label{spin_charge_rapidities}
\begin{aligned}
	&k_a= \frac{2\pi}{L}\kappa_a, && \kappa_a \in \{ 0, 1, \ldots, L-1\}, \\
	& \lambda_b = \frac{2\pi}{N} \left[ l_b - \frac{N}{2} + \frac{1}{4}\left( 1 + (-1)^{N-M} \right) \right], &&  l_b \in \{ 0, 1, \ldots, N-1 \}.
\end{aligned}
\ee
In some cases it might be convenient to shift the charge quantum numbers as $\kappa_a \to \kappa_a - \left\lfloor L/2 \right\rfloor$, where~$\lfloor x \rfloor$ denotes the integer part of~$x$.
The energy of the eigenstate~(\ref{t_0_eigenstates_fixed_NM_sector}) coincides with the one given by Eq.~(\ref{t0_spectrum}) in the main text, which reads $E({\boldsymbol k}, \varphi) = -2 t \sum_{a=1}^N \cos \left( k_a + \varphi/L\right)$.
Note that the energy only depends on the total quasimomentum $\varphi$ of the spin degrees of freedom and {\it not} on the individual spin rapidies~$\lambda_b$.

We now proceed with transforming the eigenstate~(\ref{t_0_eigenstates_fixed_NM_sector}), which belongs to the sector with fixed $N$ and $M$, into the eigenstate~(\ref{t_0_eigenstates}) that lies in the sector with a fixed spin pattern~${\bs \sigma}$. Taking into account that~${\bs \sigma}$ is conserved, it is convenient to rewrite the summation over~$\sigma_1, \ldots, \sigma_N$ in Eq.~(\ref{t_0_eigenstates}) in a different way. Namely, let us explicitly separate in Eq.~(\ref{t_0_eigenstates_fixed_NM_sector}) the summation over spin patterns~${\boldsymbol \sigma}$ that can {\it not} be transformed into each other by cyclic permutations from the summation over those patterns that {\it are} connected by a cyclic permutation:
\be \label{sum_over_spins_to_sum_over_patterns}
	\sum_{\substack{ \sigma_1 , \ldots, \sigma_N = \uparrow, \downarrow \\ \sum_{s=1}^N \delta_{\sigma_s, \downarrow}= M } }  \xi_{ {\bs \lambda} }( {\bs \sigma} )
	\prod_{j=1}^N c_{x_j, \sigma_j}^{\dag} \ket{0} = \sum_{{\boldsymbol \sigma} \in {\cal P} } \sum_{r = 0}^{{\cal Z}_{{\bs \sigma}}-1} \xi_{ {\bs \lambda} }( {\cal C}^r {\bs \sigma} )
	 \prod_{j=1}^N c_{x_j, \sigma_{j+r} }^{\dag} \ket{0},
\ee
where ${\cal P}$ is the set of inequivalent spin configurations~${\boldsymbol \sigma}$ that can not be connected by cyclic permutations, 
${\cal Z}_{{\bs \sigma}}$ is the period of pattern ${\bs \sigma}$ [see discussion after Eq.~(\ref{Hilber_space_sector_dim_calD}) in the main text],
and~${\cal C}^r {\boldsymbol \sigma}$ is the $r$-fold cyclic permutation of~${\boldsymbol \sigma} = \{ \sigma_1, \ldots, \sigma_N \}$, i.e. one has
\be \label{pattern_cyclic_perm}
	{\cal C}^r {\boldsymbol \sigma}  = \{ \sigma_{1 + r}, \ldots, \sigma_{N + r} \}.
\ee
Taking into account that the positions of down spins become  $y_m( {\cal C}^r {\bs \sigma} ) = y_m({\bs \sigma} ) - r$ [since the whole pattern is shifted by $r$ positions as shown in Eq.~(\ref{pattern_cyclic_perm})], we immediately obtain 
\be \label{spin_wavefunction_cyclic_shift}
    \xi_{{\bs \lambda}} ( {\cal C}^r {\bs \sigma} ) = \det_{1 \leq a, b \leq M} \left\{ \frac{1}{\sqrt{N}} e^{i \lambda_a y_b( {\bs \sigma} ) } e^{- i r \lambda_a} \right\} = e^{- i r \sum_{m=1}^M \lambda_a}  \xi_{{\bs \lambda}} ( {\bs \sigma} ) = e^{-i \varphi}\xi_{{\bs \lambda}} ( {\bs \sigma} ).
\ee

Thus, taking into account Eqs.~(\ref{sum_over_spins_to_sum_over_patterns}) and~(\ref{spin_wavefunction_cyclic_shift}), one can rewrite the eigenstate~(\ref{t_0_eigenstates_fixed_NM_sector})  as~\cite{Mielke1991}
\be \label{Psi_NM_to_Psi_sigma}
	 \ket{\Psi^{N,M}_{t-0}( {\boldsymbol k}, {\boldsymbol \lambda} )} 
	 \equiv \sum_{ {\boldsymbol \sigma} \in {\cal P
}} \sqrt{ {\cal Z}_{ {\boldsymbol \sigma} } } \, \xi_{{\bs \lambda}} ({\boldsymbol \lambda} ) \ket{\Psi^{{\bs \sigma}}_{t-0}( {\boldsymbol k}, \varphi )},
\ee
where the state
\begin{equation} 
	\ket{\Psi^{{\bs \sigma}}_{t-0}( {\boldsymbol k}, \varphi )} =  
 \sum_{ 1 \leq x_1 <  \ldots < x_N \leq L }  \psi_{ {\boldsymbol k}, \varphi} ({\boldsymbol x}) \, \frac{1}{ \sqrt{ {\cal Z}_{ {\boldsymbol \sigma} } } } \sum_{r=0}^{{\cal Z}_{ {\boldsymbol \sigma} }-1} e^{- i \varphi r}  \prod_{l=1}^N c_{x_l, \sigma_{l+r}}^{\dag} \ket{0}, 
\end{equation}
is nothing else than the eigenstate of the $t-0$ model in the sector with fixed spin patterns~${\bs \sigma}$, given by Eq.~(\ref{t_0_eigenstates}) in the main text. Note that the spin indices in Eq.
~(\ref{Psi_NM_to_Psi_sigma}) are periodic, $\sigma_{j+ {\cal Z}_{{\bs \sigma}} } \equiv \sigma_j$.
\section{Action of~$U_k$ in Eq.~(\ref{Kumar_unitary})}
\label{A:Unitary_transformations}

In this Appendix we derive explicit expressions for the unitary transformations $U_k^{\dag} f_j U_k$, $U_k^{\dag} \sigma_j^{\alpha} U_k$, and $U_k^{\dag} P_{i,j} U_k$, where~$U_k$ is given by Eq.~(\ref{Kumar_unitary}), $f_j$ is the fermionic annihilation operator, $\sigma_j^{\alpha}$ are the Pauli matrices, and~$P_{i,j}$ is the permutation operator from Eq.~(\ref{perm_op}). 

First of all, for the fermionic operator~$f_j$ one has

\be \label{U_on_f}
	U_k^{\dag} f_k U_k = f_k T_k, \qquad 	U_k^{\dag} f_j U_k = f_j, \quad (k\neq j),
\ee
which follows immediately from Eq.~(\ref{Kumar_unitary}) and the relation $[{\cal N}_j, f_k] = - \delta_{j,k} f_k$. Note that one clearly has~$U_k^{\dag} {\cal N}_j U_k = {\cal N}_j$ for any $j$ and $k$.

Let us now consider the transformation
\be
	U_{k}^{\dag} {\boldsymbol \sigma}_j U_{k} = (1- {\cal N}_k) {\boldsymbol \sigma}_j + {\cal N}_k T_{k}^{\dag} {\boldsymbol \sigma}_j T_{k},
\ee
where $2 \leq k \leq L$, ${\boldsymbol \sigma}_j$ is the vector of Pauli matrices, and we took into account that~${\cal N}_k^2 = {\cal N}_k$. Then, using Eq.~(\ref{T_j_action}) we obtain
\be	\label{U_k_dag_sigma_U_k}
	U_{k}^{\dag} {\boldsymbol \sigma}_j U_{k} = 
		\begin{cases}
			(1- {\cal N} _k) {\boldsymbol \sigma}_j + {\cal N}_k \, {\boldsymbol \sigma}_{j+1}, & j < k, \\
			(1- {\cal N} _j) {\boldsymbol \sigma}_j + {\cal N}_j \, {\boldsymbol \sigma}_{1}, & j=k, \\ 
			{\boldsymbol \sigma}_j, & j > k. \\
		\end{cases}
\ee

Finally, we are interested in the transformation~$U_k^{\dag} P_{i,j} U_k$.
Taking into account Eq.~(\ref{perm_op}) for the permutation operator, we have
\be
	U_k^{\dag} P_{i,j} U_k = \frac{1}{2} \left( 1 + U_k^{\dag} {\boldsymbol \sigma}_i U_k \cdot U_k^{\dag} {\boldsymbol \sigma}_j U_k \right).
\ee
Since~$P_{i,j} = P_{j,i}$, without loss of generality we may assume that~$i < j$. Then, depending on the ordering of~$(i,k)$ and  $(j,k)$, from Eq.~(\ref{U_k_dag_sigma_U_k}) one obtains three distinct cases given below:
\begin{align} 
        & U_k^{\dag} P_{i,j} U_k = \left( 1 - {\cal N}_k \right) P_{i,j} + {\cal N}_k P_{i+1, j+1},
        &&\text{($i < k$ and $j < k$)},
        \label{U_on_P_a} \\
	   &U_k^{\dag} P_{i,k} U_k = \left( 1 - {\cal N}_k \right) P_{i,k} + {\cal N}_k P_{i+1, 1},
        &&\text{($i < k$ and $j = k$)},
        \label{U_on_P_b} \\
		&U_k^{\dag} P_{i,j} U_k = \left( 1 - {\cal N}_k \right) P_{i,j} + {\cal N}_k P_{i+1, j},
        &&\text{($i < k$ and $j > k$).}
        \label{U_on_P_c}
\end{align}

\section{Derivation of~$H_B$ in Eq.~(\ref{boundary_link_hopping_v2})} \label{A:boundary_link_hopping}

In this Appendix we derive the simplified form of the boundary term
\be
    H_B - t \, {\cal U}^{\dag} P_{L,1} \left( f_{L}^{\dag} f_1 + f_1^{\dag} f_L  \right) {\cal U}.
\ee
Using the results of Appendix~\ref{A:Unitary_transformations}, it is straightforward to derive Eq.~(\ref{boundary_link_hopping_v2}). First of all, note that we can write
\be \label{boundary_link_insert_N}
	P_{1,L} f_1^{\dag} f_L = P_{1,L} {\cal N}_1 \,f_1^{\dag} f_L,
\ee
since ${\cal N}_j \,f_j^{\dag} = f_j^{\dag}$. Then, using Eq.~(\ref{Kumar_unitary}) for the operator~${\cal U}$ and taking into account Eq.~(\ref{U_on_P_c}) with $i=1$ and $j=L$, we have
\begin{multline}
	 {\cal U}^{\dag} {\cal N}_1 P_{1,L} f_1^{\dag} f_L {\cal U} = \prod_{k=L}^2 U_{k}^{\dag} {\cal N}_1 P_{1,L} f_1^{\dag} f_L  \prod_{k=2}^L U_k
	 = \prod_{k=L}^3 U_{k}^{\dag} \Bigl[ \left( 1 - {\cal N}_2 \right) {\cal N}_1 P_{1,L}
	 +  {\cal N}_2 {\cal N}_1 P_{2, L} \Bigr] f_1^{\dag} f_L \prod_{k=3}^L U_k \\
	 = \prod_{k=L}^4 U_{k}^{\dag} \Bigl[ \left( 1 - {\cal N}_3 \right) \left( 1 - {\cal N}_2 \right) {\cal N}_1 P_{1,L} 
	+ {\cal N}_3 \left( 1 - {\cal N}_2 \right) {\cal N}_1 P_{2,L}  + \left( 1 - {\cal N}_3 \right) {\cal N}_2 {\cal N}_1 P_{2, L} 
	 + {\cal N}_3 {\cal N}_2 {\cal N}_1 P_{3, L} \Bigr] f_1^{\dag} f_L \prod_{k=4}^L U_k,
\end{multline}
where the expression in the square brackets can be formally written as $P_{{\cal N}_1 + {\cal N}_2 + {\cal N}_3, \, L}$, since it produces the same result when acting on a given state in the occupation number basis. Thus, repeatedly using Eq.~(\ref{U_on_P_c}), we obtain
\be
	{\cal U}^{\dag} {\cal N}_1 P_{1,L} f_1^{\dag} f_L {\cal U} = U_L^{\dag} P_{L, \sum_{l=1}^{L-1} {\cal N}_l } f_1^{\dag} f_L U_L 
	 = \Bigl[ ( 1 -{\cal N}_L ) P_{L, \sum_{l=1}^{L-1} {\cal N}_l } 
	 + {\cal N}_L P_{1, 1 + \sum_{l=1}^{L-1} {\cal N}_l } \Bigr] f_1^{\dag} f_L T_L,
\ee
where we took into account Eq.~(\ref{U_on_f}) for $U_L^{\dag} f_L U_L$ and Eq.~(\ref{U_on_P_b}) with~$i = \sum_{l=1}^{L-1} {\cal N}_l$ and $k = L$. Then, since~${\cal N}_j f_j = 0$, one has
\begin{equation} \label{boundary_link_P_LN_TL}
	 {\cal U}^{\dag} {\cal N}_1 P_{1,L} f_1^{\dag} f_L {\cal U} = P_{L, \sum_{l=1}^{L-1} {\cal N}_l } T_L  f_1^{\dag} f_L  
	 = (1 + {\cal N}_L - {\cal N}_L) P_{L, \sum_{l=1}^{L-1} {\cal N}_l } T_L  f_1^{\dag} f_L 
	= P_{L, \sum_{l=1}^{L} {\cal N}_l } T_L  f_1^{\dag} f_L = P_{L,N} T_L f_1^{\dag} f_L,
\end{equation}	
where~$1\leq N <  L$ is the total number of fermions. Thus, using Eqs.~(\ref{boundary_link_insert_N}) and (\ref{boundary_link_P_LN_TL}), one can write the boundary hopping term~$H_B$ as
\be \label{boundary_link_hopping}
		H_{B} = - t \left(T_L^{-1}P_{L,N} \; f_L^{\dag} f_1 + P_{L,N} T_L \; f_1^{\dag} f_L \right).
\ee
Then, it is convenient to write 
\be
    P_{L,N} T_L = T_N T_L^{-N} T_{L-N} T_L^N = e^{i \Pi_N} T_L^{-N} e^{i \Pi_{L-N} } T_L^{N} = e^{i \Pi_N}  e^{i T_L^{-N} \Pi_{L-N} T_L^{N}}, 
\ee
where we took into account Eq.~(\ref{shift_op_exp_momentum}). Thus, Eq.~(\ref{boundary_link_hopping}) yields
\be 
	H_{B} = -t \left( e^{  i \, \Pi_N} e^{i \tilde \Pi_{L - N} }  f_1^{\dag} f_L + e^{ - i \tilde \Pi_{L - N} } e^{ - i \, \Pi_N}  f_L^{\dag} f_1  \right),
\ee
where $\tilde \Pi_{L - N} =  T_L^{-N} \Pi_{L - N} T_L^{N}$ and for $N=L-1$ it is understood $\Pi_1 \equiv 0$,
which gives Eq.~(\ref{boundary_link_hopping_v2}) in the main text.

\section{Derivation of~${\cal U}^{\dag} \sigma_j^{z} \, {\cal U}$ in Eq.~(\ref{random_field_transformed_H})} \label{A:sigma_transform}

Using Eq.~(\ref{Kumar_unitary}) for the unitary operator~${\cal U}$, we obtain
\begin{multline} \label{U_dag_sigma_U_1}
	{\cal U}^{\dag}  {\boldsymbol \sigma}_j {\cal U} = \prod_{k=L}^2 U_k^{\dag} \; {\boldsymbol \sigma}_j \; \prod_{k=2}^L U_k 
	= \prod_{k=L}^{j+1}U_k^{\dag} \; \Bigl[ ( 1 - {\cal N}_j ) {\boldsymbol \sigma}_j + {\cal N}_j {\boldsymbol \sigma}_1 \Bigr]\; \prod_{k=j+1}^L U_{k}\\
	= \prod_{k=L}^{j+2}U_k^{\dag} \; \Bigl[ ( 1 - {\cal N}_j )  \Bigl( ( 1 - {\cal N}_{j+1} ){\boldsymbol \sigma}_j
	 + {\cal N}_{j+1} {\boldsymbol \sigma}_{j+1} \Bigr)
	 + {\cal N}_j \Bigl( ( 1 - {\cal N}_{j+1} ) {\boldsymbol \sigma}_1 + {\cal N}_{j+1} {\boldsymbol \sigma}_2 \Bigr)  \Bigr]\; \prod_{k=j+2}^L U_{k},
\end{multline}
where we assumed that~$2 \leq  j \leq L-1$ and took into account Eq.~(\ref{U_k_dag_sigma_U_k}). We now observe that the expression in the square brackets can be formally written as
\be
	 (1- {\cal N}_j) {\boldsymbol \sigma}_{j + {\cal N}_{j + 1}} +  {\cal N}_j {\boldsymbol \sigma}_{1 + {\cal N}_{j + 1}},
\ee
since it produces the same result when acting on a given state in the occupation number basis.
Continuing the procedure, from Eq.~(\ref{U_dag_sigma_U_1}) we obtain
\be
	{\cal U}^{\dag}  {\boldsymbol \sigma}_j {\cal U} = \prod_{k=L}^{L-p}U_k^{\dag}  \Bigl[  (1 - {\cal N}_j) {\boldsymbol \sigma}_{ j + \sum_{l=j+1}^{L-p-1}{\cal N}_{l} } 
	  + {\cal N}_j {\boldsymbol \sigma}_{ 1 + \sum_{l=j+1}^{L-p-1}{\cal N}_{l} } \Bigr] \prod_{k=L-p}^L U_{k}, 
\ee
where $1 \leq p \leq L - j -1$. Finally, we arrive at the relation
\be \label{U_dag_sigma_j_U_2}
	{\cal U}^{\dag}  {\boldsymbol \sigma}_j {\cal U} = (1 - {\cal N}_j) {\boldsymbol \sigma}_{j + \sum_{k=j+1}^L {\cal N}_{k} } 
	 + {\cal N}_j {\boldsymbol \sigma}_{1 + \sum_{k=j+1}^L {\cal N}_{k} }, 
\ee
where $1<j<L$.
The cases $j=1$ and $j=L$ should be treated separately, and we obtain
\be \label{U_dag_sigma_1_L_U_2}
	{\cal U}^{\dag} {\boldsymbol \sigma}_1 {\cal U} = {\boldsymbol \sigma}_{1 + \sum_{k=2}^L {\cal N}_k }, \qquad \qquad 
	{\cal U}^{\dag} {\boldsymbol \sigma}_L {\cal U} = (1 -  {\cal N}_L) {\boldsymbol \sigma}_{L} + {\cal N}_L {\boldsymbol \sigma}_{1}.
\ee
From Eqs.~(\ref{U_dag_sigma_j_U_2}) and~(\ref{U_dag_sigma_1_L_U_2}) we see that the transformation~(\ref{Kumar_unitary}) shifts ${\boldsymbol \sigma}_j$ to some other position $j^{\prime}$ that depends on the number of occupied sites in the interval $[j + 1, L]$ of the chain. In other words, ${\cal U}^{\dag} {\boldsymbol \sigma}_j {\cal U}$ effectively acts as a permutation. It then follows immediately that
$
	{\cal U}^{\dag}\; \sum_{j=1}^L {\boldsymbol \sigma}_j \; {\cal U} = \sum_{j=1}^L {\boldsymbol \sigma}_j.
$

\section{Overlaps involving Slater determinants} \label{A:Slater_dets}

For completeness, in this Appendix we present detailed calculations of various overlaps involving Slater determinants (see e.g. Ref.~\cite{Lowdin1955} for a related discussion). Consider the following Slater determinant of single particle states~$\psi_{n } (x)$:
\be
	\psi( {\bs n}, {\bs x} ) = \det_{1\leq a,b \leq N} \left\{ \psi_{n_a}(x_b) \right\} = \sum_{ P \in {\cal S}_N } (-1)^{P} \prod_{j=1}^N \psi_{n_j} (x_{Pj}) = \sum_{ P \in {\cal S}_N } (-1)^{P} \prod_{j=1}^N \psi_{n_{Pj} } (x_{j}) \equiv {\cal A} \prod_{j=1}^N \psi_{n_{j} } (x_{j}),
\ee
where~${\cal S}_N$ is the symmetric group of order~$N$,~$P$ is a permutation~$\{1, 2, \ldots, N \} \to \{ P1, P2, \ldots, PN \}$, and  we introduced the antisymmetrization operator~${\cal A}$, which is Hermitian and satisfies~${\cal A}^2 = N! \, {\cal A}$. 

\subsection{Inner product of Slater determinants}

For the inner product of two Slater determinants one has
\be  \label{Slater_dets_overlap}
\begin{split}
	 \left\langle \phi( {\bs m} ) \right| \left. \! \psi({\bs n}) \right\rangle  &\equiv \frac{1}{N!}\sum_{x_1 = 1}^L \! \ldots \!  \sum_{x_N = 1}^L \phi^*( {\bs m}, {\bs x} ) \psi( {\bs n}, {\bs x} ) 
	  = \frac{1}{N!} \sum_{x_1 = 1}^L \! \ldots \!  \sum_{x_N = 1}^L \sum_{P \in {\cal S}_N } (-1)^P \prod_{j=1}^N \phi^*_{m_j}(x_{Pj}) {\cal A} \prod_{j=1}^N \psi_{n_j}(x_j) \\
	&=\frac{1}{N!} \sum_{x_1 = 1}^L \! \ldots \!  \sum_{x_N = 1}^L \prod_{j=1}^N \phi^*_{m_j}(x_{j}) \sum_{P \in {\cal S}_N } (-1)^P {\cal A} \prod_{j=1}^N \psi_{n_j}(x_{Pj}) =  \sum_{P \in {\cal S}_N } (-1)^P \prod_{j=1}^N \sum_{x = 1}^L \phi^*_{m_j}(x) \psi_{n_{Pj}}(x) \\ 
	 &= \det_{1\leq a,b \leq b} \left\{ \sum_{x = 1}^L \phi^*_{m_a}(x) \psi_{n_b}(x) \right\},
\end{split}
\ee
which is nothing else than the determinant of single particle inner products. For orthonormal single particle states satisfying~$\sum_{x = 1}^L \phi^*_{m_a}(x) \psi_{n_b}(x) = \delta_{m_a, n_b}$ we arrive at~$\left\langle \phi( {\bs m} ) \right| \left. \! \psi({\bs n}) \right\rangle = \delta_{{\bs m}, {\bs n}}$.

\subsection{ Matrix elements of a symmetric single-body operator~$V_1^{\text{sym}}(x_1, \ldots, x_N) = \sum_{l=1}^N f(x_l)$}

Consider a single-body operator~$V_1^{\text{sym}}({\bs x}) = \sum_{l = 1}^N f (x_l)$, which is totally symmetric under an arbitrary exchange of the coordinates~$x_1, \ldots, x_N$. Its matrix element between two Slater determinants are given by
\be \label{Slater_dets_ME_sym_1}
\begin{split} 
	 \bra{ \phi( {\bs m} ) } V_1^{\text{sym}} \ket{ \psi({\bs n})}  &= \frac{1}{N!} \sum_{x_1 = 1}^L \! \ldots \!  \sum_{x_N = 1}^L \phi^*( {\bs m}, {\bs x} ) \sum_{l = 1}^N f( x_l ) \psi( {\bs n}, {\bs x} ) \\ 
	 &= \frac{1}{N!} \sum_{x_1 = 1}^L \! \ldots \!  \sum_{x_N = 1}^L \sum_{P \in {\cal S}_N } (-1)^P \prod_{j=1}^N \phi^*_{m_j}(x_{Pj}) \sum_{l = 1}^N f( x_{l} ) {\cal A} \prod_{j=1}^N \psi_{n_j}(x_j) \\
	&= \frac{1}{N!}\sum_{x_1 = 1}^L \! \ldots \!  \sum_{x_N = 1}^L \prod_{j=1}^N \phi^*_{m_j}(x_{j}) \sum_{l = 1}^N f( x_{l} ) \sum_{P \in {\cal S}_N } (-1)^P {\cal A} \prod_{j=1}^N \psi_{n_j}(x_{Pj})  \\ 
	 &= \sum_{l=1}^N \sum_{s=1}^N \; \sum_{x =1}^L f(x)\phi^*_{m_l}(x) \psi_{n_s}(x) \sum_{Q \in {\cal S}_N } (-1)^Q \delta_{s, Ql} \prod_{ \substack{ j=1 \\ (j \neq l)} }^N \sum_{x = 1}^L \phi^*_{m_j}(x) \psi_{n_{Qj}}(x)\\
	 &= \sum_{l=1 }^N \sum_{s=1 }^N (-1)^{l+s} \sum_{x =1}^L f(x)\phi^*_{m_l}(x) \psi_{n_s}(x) \det_{ \substack{1\leq a,b \leq N\\ (a \neq l, b \neq s)} } \left\{ \sum_{x = 1}^L \phi^*_{m_a}(x) \psi_{n_b}(x) \right\},
\end{split}
\ee
where in the last equality the factor of $(-1)^{l+s}$ appears after the summation over $Q$ because of the Kronecker symbol~$\delta_{s,Ql}$. Setting $f(x) = 1/N$ in Eq.~(\ref{Slater_dets_ME_sym_1}), so that one has $V_1^{\text{sym}} \equiv 1$, we immediately see that one of the summations (say the one over $l$) gives the cofactor expansion of the determinant. Then, the remaining sum (say, the one over $s$) has $N$ identical terms, so that Eq.~(\ref{Slater_dets_ME_sym_1}) reduces to~Eq.~(\ref{Slater_dets_overlap}).

\subsection{ Matrix elements of a symmetric many-body operator~$V^{\text{sym}}(x_1, \ldots, x_N) = \sum_{l=1}^N e^{i \theta \nu_{1, x_{l}} }f(x_{l})$} \label{Slater_dets_ME_sym_many_body}

Finally, we consider a many-body operator $V^{\text{sym}}(x_1, \ldots, x_N) = \sum_{l=1}^N e^{i \theta \nu_{ 1, x_{l} }} f(x_{l})$, where $\nu_{ 1, x_{l} }$ is the counting function given by Eqs.~(\ref{counting_function_definition}) and (\ref{counting_function_action}) in the main text. Taking into account that for the ordered sector $x_{1} < \ldots < x_l < \ldots < x_{N}$ one has $\nu_{1, x_{l}} = l$, we write 
\begin{multline} \label{Slater_dets_ME_sym_MB}
    \bra{ \phi( {\bs m} ) } V^{\text{sym}} \ket{ \psi({\bs n})}  = \frac{1}{N!} \sum_{x_1 = 1}^L \! \ldots \!  \sum_{x_N = 1}^L \phi^*( {\bs m}, {\bs x} ) \sum_{r = 1}^N e^{i \theta \nu_{ 1, x_{r} } } f( x_{r} ) \psi( {\bs n}, {\bs x} ) \\
     = \sum_{r=1}^N e^{i \theta r } \sum_{1 \leq x_1 < \ldots < x_N \leq L} f(x_{r}) \sum_{P,Q \in {\cal S}_N} (-1)^{P+Q}\prod_{j=1}^N \phi^*_{m_{Pj}}(x_{j}) \psi_{n_{Qj}}(x_{j}) ,
\end{multline}
where we replaced the repeated sumation over $x_1, \ldots x_N$ with the ordered one. We then rewrite the ordered summation in the second line of Eq.~(\ref{Slater_dets_ME_sym_MB}) as follows
\be \label{ordered_sum_splitting}
    \sum_{1 \leq x_1 < \ldots < x_N \leq L} \to \left( \sum_{x_1 = 1}^{x_2} \ldots \sum_{x_{r-1} = 1}^{x_{r}} \right)\; \sum_{x_{r} = 1}^L \; \left( \sum_{ x_{r + 1} = x_{r} + 1 }^L \ldots \sum_{x_N = x_{N-1} + 1}^L \right).
\ee
Note that in Eq.~(\ref{ordered_sum_splitting}) we extended the ordered sector to $1 \leq x_1 \leq \ldots \leq x_{r-1} \leq x_r < \ldots < x_N$. This is allowed since the terms with $x_{j-1} = x_{j}$ do not contribute to the matrix element~(\ref{Slater_dets_ME_sym_MB}) due to the presence of Slater determinants.
Thus, using Eq.~(\ref{ordered_sum_splitting}) we reduce Eq.~(\ref{Slater_dets_ME_sym_MB}) to
\begin{multline} \label{Slater_dets_ME_sym_MB_2}
    \bra{ \phi( {\bs m} ) } V^{\text{sym}} \ket{ \psi({\bs n})}  = \sum_{r=1}^N e^{i \theta r } \sum_{P,Q \in {\cal S}_N} (-1)^{P+Q} \sum_{x_{r} =1 }^L f(x_{r}) \phi^*_{m_{P r}}(x_{r}) \psi_{n_{Q r}}(x_{r}) \\
     \times \sum_{x_1 = 1}^{x_2} \ldots \sum_{x_{r-1} = 1}^{x_{r}} 
    \prod_{j=1}^{r-1} \phi^*_{m_{Pj}}(x_{j}) \psi_{n_{Qj}}(x_{j}) 
    \sum_{ x_{r + 1} = x_{r} + 1 }^L \! \ldots \! \sum_{x_N = x_{N-1} + 1}^L 
    \prod_{j=r+1}^N \phi^*_{m_{Pj}}(x_{j}) \psi_{n_{Qj}}(x_{j}).
\end{multline}
Note that because of the summation over the permutations $P$ and $Q$, the summands in the ordered summations over $x_1 \leq \ldots \leq x_{r - 1}$ and $x_{r + 1} < \ldots < x_N$ in the second line of Eq.~(\ref{Slater_dets_ME_sym_MB_2}) are symmetric functions of the corresponding variables. This allows us to replace the ordered summations with the repeated ones:
\be
    \sum_{1 \leq x_1 \leq \ldots \leq x_{r-1} \leq x_{r}} \to \frac{1}{(r-1)!} \sum_{x_1 = 1}^{x_{r}} \! \ldots \! \sum_{x_{r-1} = 1}^{x_{r}}, 
    \qquad 
    \sum_{x_{r}+1 \leq x_{r+1} < \ldots < x_{N} \leq L } \to \frac{1}{(N-r)!} \sum_{x_{r+1} = x_{r}+1}^L \! \ldots \! \sum_{x_{N} = x_{r}+1}^L.
\ee
Then Eq.~(\ref{Slater_dets_ME_sym_MB_2}) can be written as
\begin{multline} \label{Slater_dets_ME_sym_MB_3}
    \bra{ \phi( {\bs m} ) } V^{\text{sym}} \ket{ \psi({\bs n})}  = \sum_{l=1}^N \sum_{s=1}^N  \sum_{x = 1}^L f(x) \phi^*_{m_{l}}(x) \psi_{n_{s}}(x) \\
     \times \sum_{r=1}^N \frac{ e^{i \theta r } }{(N-r)!(r-1)!} 
    \sum_{ P,Q \in {\cal S}_N } (-1)^{Q+P} \delta_{l, P r} \delta_{s, Q r}\prod_{j=1}^{r-1} \sum_{y=1}^{x} \phi^*_{m_{Pj}}(y) \psi_{n_{Qj}}(y) 
    \prod_{j=r+1}^N \sum_{y=x+1}^L \phi^*_{m_{Pj}}(y) \psi_{n_{Qj}}(y).
\end{multline}   
One can easily check that Eq.~(\ref{Slater_dets_ME_sym_MB_3}) significantly simplifies and it yields
\begin{multline}
    \bra{ \phi( {\bs m} ) } V^{\text{sym}} \ket{ \psi({\bs n})}  = e^{i \frac{N+1}{2} \theta } \sum_{l=1}^N \sum_{s=1}^N (-1)^{l+s}  \sum_{x = 1}^L  f(x) \phi^*_{m_{l}}(x) \psi_{n_{s}}(x)\\
    \times \det_{\substack{1\leq a,b \leq N \\ (a\neq l, b \neq s)}} 
    \left\{ \sum_{y=1}^L e^{i \,\text{sgn}(x-y) \theta /2} \phi^*_{m_{a}}(y) \psi_{n_{b}}(y)
    \right\},
\end{multline}
where we used the fact that one can write $e^{i \theta /2} \sum_{y = 1}^{x} \phi^*_{m_{a}}(y) \psi_{n_{b}}(y) + e^{- i \theta /2} \sum_{y = x+ 1}^{L} \phi^*_{m_{a}}(y) \psi_{n_{b}}(y) = \sum_{y=1}^L e^{i \,\text{sgn}(x-y) \theta /2} \phi^*_{m_{a}}(y) \psi_{n_{b}}(y)$, since the term with $y=x$ gives zero contribution to the matrix element~(\ref{Slater_dets_ME_sym_MB_3}).

\section{Special cases of the matrix elements~(\ref{random_field_matrix_element_3})} \label{app_ME_special_cases}

\subsection{Matrix elements with $\varphi^{\prime} = \varphi$}

First, we consider matrix elements that are diagonal in the spin degree of freedom, i.e. those with $\varphi = \varphi^{\prime}$. 
From Eq.~(\ref{S_gen_def}) one immediately sees that we have
\be \label{random_field_ME_G_diag}
	S^{ {\bs \sigma} }_0 = 1 - \frac{2 M}{N}.
\ee 
Using the results of Appendix~\ref{A:Slater_dets}, in particular Eq.~(\ref{Slater_dets_ME_sym_1}), 
we can further simplify the charge part~(\ref{F_gen_def}), which in the case of~$\varphi^{\prime} = \varphi$ yields

\begin{equation} \label{F_phi_diagonal}
	F_{ \varphi , \varphi } ({\bs r}, {\bs r}^{\prime} ) = \sum_{l=1 }^N \sum_{s=1 }^N (-1)^{l+s}  \prod_{\substack{a=1 \\ a\neq l}}^N \prod_{\substack{b=1 \\ b\neq s}}^N \delta_{r_a, r^{\prime}_b} 
	 \sum_{x =1}^L h_x \psi_{\varphi}^*( r_l, x ) \psi_{\varphi}( r_s^{\prime}, x ) ,
\end{equation}
where we used the fact that the single particle wave functions with the same value of~$\varphi$ are orthonormal, $\sum_{x=1}^L \psi_{\varphi}^*( r_a, x ) \psi_{\varphi}( r_b^{\prime}, x ) = \delta_{r_a, r_b^{\prime}}$.


It is easy to see that in the case of a single spin component, i.e. for~$M=0$ or $M=N$ one has~${\cal Z}_{ {\boldsymbol \sigma} } = 1$ so that~$\varphi = \varphi^{\prime} = 0$ is the only possibility. Therefore, in the single-component case we obtain~$S^{ {\bs \sigma} }_0 = 1$ for~$M=0$ and $S^{ {\bs \sigma} }_0 = -1$ for $M=N$. The charge part of the matrix element in this case follows from Eq.~(\ref{F_phi_diagonal}) with~$\varphi = \varphi^{\prime} = 0$.

\subsection{N\'eel spin configuration}
We now look at a less trivial case, namely the N\'eel spin configuration. In this case we have~$M=N/2$, ${\cal Z}_{ {\bs \sigma} } = 2$ (recall that ${\cal Z}_{{\bs \sigma}}$ coincides with the period of a spin pattern), and~$\varphi, \varphi^{\prime} \in \{ 0, \pi \}$ [see Eq.~(\ref{t0_quantization})], so that we have $\varphi- \varphi^{\prime} \in \{0, \pm \pi\}$. 
Consider first $\varphi = \varphi^{\prime}$. In this case from Eq.~(\ref{random_field_ME_G_diag}) we have~$S^{ {\bs \sigma} }_0 = 0$, so that the diagonal in spin ($\varphi = \varphi^{\prime}$) matrix elements~(\ref{random_field_matrix_element_3}) vanish. Without loss of generality we may assume that the spin pattern~${\boldsymbol \sigma}$ always has~$\sigma_1 = \, \Downarrow$, so that~$y_1 = 1$. Then, from Eq.~(\ref{S_gen_def}) we see that for the off-diagonal~($\varphi^{\prime} \neq \varphi$) matrix elements the spin part is simply~$S^{ {\boldsymbol \sigma} }_{\pm \pi} = 1$, and Eq.~(\ref{random_field_matrix_element_3}) yields

\begin{equation}
        	\langle \Psi^{{\bs \sigma}} ( {\bs r}, \varphi ) | \tilde V_{\text{spin}} | \Psi^{{\bs \sigma} } ( {\bs r}^{\prime}, \varphi^{\prime} ) \rangle 
	= 
    \left( 1 - \delta_{ \varphi, \varphi^{\prime} } \right)
    F_{ \varphi , \varphi^{\prime}  } ( {\bs r}, {\bs r}^{\prime} ).
\end{equation}

\end{widetext}


\begin{thebibliography}{99}
\bibitem{counting_function_example}
    For instance, for ${\bs x} = \{ x_1 = 2, x_2 = 3, x_3 = 1\}$ one has $\nu_{1,x_1} \ket{{\bs x}} = 2 \ket{{\bs x}}$, $\nu_{1,x_2} \ket{{\bs x}} = 3 \ket{{\bs x}}$, and $\nu_{1,x_3} \ket{{\bs x}} = \ket{{\bs x}}$.
\end{thebibliography}
\end{document}